\documentclass[journal]{IEEEtran}
\IEEEoverridecommandlockouts
\usepackage{cite}
\usepackage[dvips]{graphicx}
\usepackage[cmex10]{amsmath}
\usepackage{algorithmic}
\usepackage{algorithm}
\usepackage{array}
\usepackage{mdwmath}
\usepackage{amssymb}
\usepackage{color}
\usepackage{subfigure}
\newtheorem{prop}{Proposition}
\newtheorem{theorem}{Theorem}
\newtheorem{lemma}{Lemma}

\begin{document}
\title{Robust Beamforming Design for Intelligent Reflecting Surface Aided Cognitive Radio Systems with Imperfect Cascaded CSI}
\author{Lei~Zhang,~\IEEEmembership{Member,~IEEE,} Cunhua~Pan,~\IEEEmembership{Member,~IEEE,} Yu~Wang,~\IEEEmembership{Member,~IEEE,} Hong~Ren,~\IEEEmembership{Member,~IEEE,} and~Kezhi~Wang,~\IEEEmembership{Member,~IEEE}
\thanks{This work was supported in part by the National Natural Science Foundation of China under Grants 61701202 and 61901196, in part by the Open Research Fund of National Mobile Communications Research Laboratory, Southeast University, under Grants 2019D17 and 2021D14, in part by the Natural Science Foundation of the Higher Education Institutions of Jiangsu Province under Grant 19KJB510026, in part by the National Key Research and Development Project under Grant 2019YFE0123600, and in part by the Jiangsu Overseas Visiting Scholar Program for University Prominent Young and Middle-aged Teachers and Presidents. \emph{(Corresponding author: Cunhua Pan.)}}
\thanks{Lei Zhang, Yu Wang, are with the School of Electrical and Information Engineering, Jiangsu University of Technology, Changzhou,
Jiangsu, 213001 China, and with National Mobile Communications Research Laboratory, Southeast University, Nanjing 210096, China, e-mail: \{zhlei, yuwang\_edina\}@jsut.edu.cn.}
\thanks{Cunhua Pan is with the School of Electronic Engineering and Computer Science, Queen Mary University of London, London E1 4NS, U.K., e-mail: c.pan@qmul.ac.uk.}
\thanks{Hong Ren is with National Mobile Communications Research Laboratory, Southeast University, Nanjing, China, e-mail: hren@seu.edu.cn.}
\thanks{Kezhi Wang is with Department of Computer and Information Sciences, Northumbria University, UK, e-mail: kezhi.wang@northumbria.ac.uk.}
}

\maketitle
\begin{abstract}
In this paper, intelligent reflecting surface (IRS) is introduced to enhance the network performance of cognitive radio (CR) systems. Specifically, we investigate robust beamforming design based on both bounded channel state information (CSI) error model and statistical CSI error model for primary user (PU)-related channels in IRS-aided CR systems. We jointly optimize the transmit precoding (TPC) at the secondary user (SU) transmitter (ST) and phase shifts at the IRS to minimize the ST's total transmit power subject to the quality of service of SUs, the limited interference imposed on the PU and unit-modulus of the reflective beamforming. The successive convex approximation (SCA) method, Schur's complement, General sign-definiteness principle, inverse Chi-square distribution and penalty convex-concave procedure are invoked for dealing with these intricate constraints. The non-convex optimization problems are transformed into several convex subproblems and efficient algorithms are proposed. Simulation results verify the efficiency of the proposed algorithms and reveal the impacts of CSI uncertainties on ST's minimum transmit power and feasibility rate of the optimization problems. Simulation results also show that the number of transmit antennas at the ST and the number of phase shifts at the IRS should be carefully chosen to balance the channel realization feasibility rate and the total transmit power.
\end{abstract}

\IEEEpeerreviewmaketitle

\section{Introduction}
As a revolutionary technique, intelligent reflecting surface (IRS) has received extensive attention from both academia and industry since it can enhance both the spectral and energy efficiency of the wireless communication systems through a preprogrammed controller\cite{Wu2019Towards,Huang-2019-Recon,Pan-2019-MultiMIMO,Huang2020Reconfi,PanRIS6G}. IRS is equipped with a large number of elements made of special materials and functioned by adjusting the reflecting coefficients (i.e., phase or amplitude) of the incident radio-frequency (RF) wave and reflecting it passively. The signals reflected by IRS can be added with other signal paths either to increase the signal strength at the desired receiver, or to mitigate the co-channel interference at the unintended users. The existing contributions have demonstrated benefits brought by introducing an IRS into wireless communication systems. For instance, some certain performance metrics such as channel capacity \cite{Pan-2019-MultiMIMO,Wu-2019-MIMO,Pan-2019-SWIPT,Zhang-2019-CaMIMO,Zhou2019Intel,Huang-2019-Recon,Han2019Large}, physical layer security rate \cite{Chen-2019-Intel,Cui-2019-Secur,Shen-2019-Secre,Yu-2019-Enabl}, transmission latency and total transmit power \cite{Bai2019Latency,Gui2019Robust} are efficiently enhanced by jointly optimizing the active transmit precoding (TPC) at the base station (BS) and the passive beamforming at the IRS.
\par Another effective technology to enhance spectrum efficiency is cognitive radio (CR) \cite{Zhang-2018Cog,Xuemin2014Cog,A12017Wireless,Tseng-2015Ul,Hamza2015Throughput,Zhao2008Opportunistic}. In CR systems, the primary user (PU) is defined as a spectrum-licensed user who always has high priority to access the spectrum, while the secondary user (SU) is normally unlicensed but can be allowed to share the spectrum without causing harmful interference to the PU. However, the challenge of CR systems is that the performance improvements for the PU and the SU are conflicting \cite{Xuemin2014Cog,A12017Wireless,LiuYuanSecure,Hamza2015Throughput,Zhao2008Opportunistic}. In specific, increasing the transmit power at the SU's BS to enhance the signal strength will generate increased interference towards the PU. Fortunately, this bottleneck can be addressed by introducing an IRS into a CR system thanks to the IRS's reconfiguration function that can help enhance the desired signal strength of the SU and mitigate the co-channel interference to the PU through jointly optimizing the TPC and passive beamforming \cite{Lei2020Intelligent,Makarfi2020Reconfi,2020HeRecon}.
\par However, the above papers \cite{Lei2020Intelligent,Makarfi2020Reconfi,2020HeRecon} studied the transmission design based on perfect channel state information (CSI), which are challenging to realize in practice. The reason is that the channels between SUs and PUs are more difficult to estimate due to the uncooperative relationship between them. The channel estimation error is inevitable. For some users or devices which require strict quality of service (QoS) requirements, it is imperative to study the robust beamforming design for the IRS-aided CR systems to guarantee each user's QoS under arbitrary channel estimation error. In IRS-aided CR system, the sensed channels related to the PU can be divided into two branches. The first one is the direct channel spanning from the SU's BS to the PU (BS-PU). The second one consists of two IRS-related channels spanning from the SU's BS to the IRS (BS-IRS) and the IRS to the PU (IRS-PU). In fact, in general IRS-aided systems without CR, there were some contributions on the robust beamforming design based on the assumption that only the channel from the IRS to the user (IRS-user) was imperfectly estimated \cite{Zhang-2019-CaMIMO,Gui2019Robust,Yu2019Robust}. However, it is very challenging to estimate the BS-IRS channel and IRS-user channel independently since the IRS is passive and can neither send nor receive pilot symbols. Installing some active elements at the IRS will increase an undesired burden of the IRS due to the increased hardware and extra power cost. Additionally, the extra information exchange overhead is required to feed back from the IRS to the BS. Therefore, another approach to design the robust beamforming is based on the assumption that the two IRS-related channels are regarded as a cascaded BS-IRS-user channel, which is the product of the BS-IRS channel and the IRS-user channel. It is more cost-effective to estimate the cascaded BS-IRS-user channel since no active RF chains are required at the IRS. It has been verified that considering cascaded BS-IRS-user channel is sufficient for the beamforming design \cite{Zhou2019Joint,Wang2019Channel,Wang2019Compressed,Chen2019Channel}. Based on the cascaded channel, there are some contributions on robust beamforming design in IRS-aided systems \cite{Zhou2020Framework}. In \cite{Zhou2020Framework}, a framework of robust transmission beamforming was proposed based on imperfect cascaded IRS-related channels at the transmitter. The worst-case and outage probability robust beamforming designs were provided by minimizing the total transmit power. However, the above robust beamforming designs are not applicable for the CR network since the architecture and optimization problem of IRS-aided CR systems are completely different from traditional CR systems.
\par To the best of our knowledge, only a few contributions studied transmission beamforming design based on imperfect PU-related channels for IRS-aided CR systems \cite{Xu2020ResourceJ,Yuan2019Intel}. The authors in \cite{Xu2020ResourceJ} studied the robust beamforming design under bounded CSI uncertainty of PU-related channels in IRS-aided full-duplex CR systems with the aim of maximizing the system sum rate of SUs. However, they considered the imperfect IRS-PU channel instead of the imperfect cascaded BS-IRS-PU channel. In \cite{Yuan2019Intel}, the cascaded BS-IRS-PU channel was first considered to be imperfect with the assumption of the bounded CSI error model, based on which the authors proposed a robust beamforming design to maximize the single user's data rate. The simulation results therein showed that the SU's achievable rate is significantly improved in an IRS-aided multi-input single-output (MISO) CR system compared with a CR system without IRS. However, the above mentioned contributions that targeted at maximizing the capacity cannot guarantee the quality of service (QoS) requirements of each SU, which cannot be applied in some emerging applications with stringent QoS requirements such as video conference, autonomous vehicles, etc.
\par Against the above background, in this paper, we investigate robust beamforming design aiming at minimizing the total transmit power of SUs subject to each SU's QoS requirement and the interference limit imposed on the PU. The imperfect PU-related cascaded channel is considered to avoid the requirement of additional RF chains at the IRS. In contrast to the rate maximization problem in \cite{Xu2020ResourceJ} and \cite{Yuan2019Intel}, the power minimization problem may be infeasible due to the conflicting constraints of SU's QoS requirements and PU's limited interference imposed by the SU. Different from \cite{Xu2020ResourceJ} and \cite{Yuan2019Intel} where only the bounded CSI error model is considered, in this paper, we additionally consider another type of CSI error, i.e., statistical CSI error, which is closely relevant to the channel estimation error. Specifically, the contributions of this paper can be summarized as follows:
\begin{itemize}
  \item The robust beamforming design under both the bounded and statistical CSI error models are studied for the IRS-aided CR system. The TPC matrix at the SU transmitter (ST) and the reflective element diagonalized (RED) matrix of the IRS are jointly optimized to minimize the total transmit power at the ST subject to the unit-modulus constraint of reflective elements, the QoS requirement of each SU receiver (SR) and the interference limit imposed on the PU receiver (PR). The power minimization problem may be infeasible due to the conflicting constraints of SU's QoS requirements and PU's limited interference imposed by the SU. This motivate us to check the feasibility of optimization problems and analyze the feasibility for each channel realization. The block coordinate descent (BCD) algorithm is employed to alternately optimize the TPC matrix and the RED matrix. Meanwhile, the feasibility checking for optimization problems is analyzed.
  \item For the bounded CSI error model, we consider the worst-case transmission beamforming design, where the channel error is characterized by two separated PU-related channels. In order to deal with these tricky constraints, we develop artful mathematical derivations. The successive convex approximation (SCA) method, Schur's complement, General sign-definiteness principle and penalty CCP method are adopted to transform the non-convex problems into second-order cone programming (SOCP) problems. This scheme is named as SCD scheme since which is based on the separating cascaded channel and direct channel.
  \item For the statistical CSI error model, the CSI error follows the circularly symmetric complex Gaussian (CSCG) distribution. This scheme named as STA also aims to minimize the total transmit power at the ST. The inverse Chi-square distribution is used to simplify the probabilistic constraint of the interference limit imposed on the PR. Finally, the problem for optimizing TPC matrix is transformed into an SDP problem and the problem for optimizing RED matrix is transformed into an SOCP problem.
  \item Some important results are obtained. With the assistance of the IRS, the number of phase shifts should be carefully chosen to obtain a tradeoff between the total minimum transmit power and the feasibility rate of the optimization problem. Moreover, improving the uncertainty level of PU-related channels can reduce the ST's transmit power, while a high uncertainty level will lead to a low probability to find the optimal beamforming.
\end{itemize}
\par The remainder of this paper is organized as follows. Section \ref{system model} describes the system model and gives the problem formulation. Sections \ref{section_SCD} and \ref{STA} provide the SCD robust design based on bounded error model and the STA robust design based on statistical error model, respectively. The feasibility checking problems are analyzed in Section \ref{section6}. Section \ref{section5-} researches multiple PRs scenario. Section \ref{section5--} shows the simulation results. Finally, we conclude this paper in Section \ref{section5}.
\par \emph{Notations}: The symbols such as $\mathbf{A}$ and $\mathbf{a}$ are complex-valued matrix and vector, respectively. $\mathbb{C}$ and $\mathbb{C}^{a\times b}$ are complex value set and space of $a\times b$ complex-valued matrix, respectively. $\mathrm{diag}\{\cdot\}$ is diagonalization operation. $\mathbb{E}[\cdot]$ represents mathematical expectation. $\mathbf{A}^*$, $\mathbf{A}^\mathrm{T}$ and $\mathbf{A}^\mathrm{H}$ mean the conjugate, transpose and Hermitian of $\mathbf{A}$. $\mathcal{CN}(0,\sigma)$ is the CSCG distribution with zero mean and variance $\sigma$. $|\cdot|$, $\|\cdot\|_2$ and $\|\cdot\|_\mathrm{F}$ represent modulo, Euclidean norm and Frobenius norm, respectively. $\mathbf{I}_a$ is $a\times a$ unitary matrix. $\mathrm{vec}(\mathbf{A})$, $\mathrm{Tr}(\mathbf{A})$ and $\mathrm{Re}(\mathbf{A})$ mean vectorization, trace and the real part of $\mathbf{A}$, respectively. For the sake of readability, the abbreviations are listed in Table \ref{tab1}.
\begin{table}\centering
\caption{Explanations of abbreviations.}
\label{tab1}
\begin{tabular}{|c|c|}
  \hline
  Abbreviations & Explanations\\
  \hline
  TPC & Transmit precoding \\
  RED & Reflective element diagonalize \\
  ST & Secondary user's transmitter \\
  SR & Secondary user's receiver \\
  PR  & Primary user's receiver\\
  BCD & Block coordinate descent \\
  SCD & Separating cascaded channel and direct channel\\
  SDP & Semi-definite programming\\
  CCP & Convex-concave procedure \\
  STA & Statistical CSI error\\
  IT & Interference temperature \\
  SOCP &Second-order cone programming\\
  \hline
\end{tabular}
\end{table}
\section{System Model and Problem Formulation}\label{system model}
\begin{figure}[!t]
\centering
\includegraphics[width=2.4in]{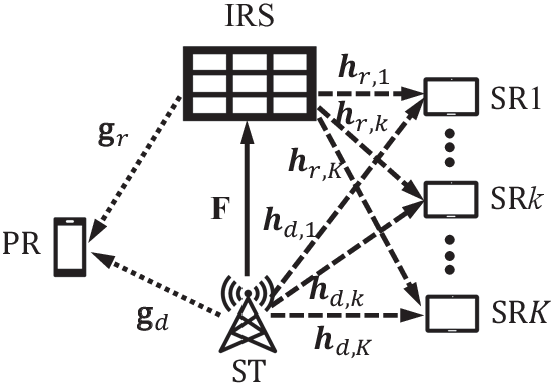}
\caption{IRS-aided CR.}
\label{fig1}
\end{figure}
\subsection{Signal Transmission Model}
In order to improve the performance of SUs, an IRS is deployed in the CR system shown in Fig. \ref{fig1} where we consider the downlink MISO transmission. The system consists of one IRS, one PR, one ST and $K$ SRs, where the superimposed signals of $K$ SUs are transmitted from the ST. The ST is equipped with $M_t$ transmit antennas and the PR (or each SR) is equipped with a single receive antenna. Each SR will receive the encoded signals via the IRS or directly from the ST and then decode its own signal. Similarly, the PR will also receive the interference signals via the IRS or from the ST directly. The IRS which is equipped with $N$ reflective elements can receive the transmitted signals and passively reflect them without additional RF electric circuit. Each reflective element is denoted by $\phi_n=e^{j\theta_n}, n\in\mathcal{N}=\{1, 2, \cdots, N\}$\footnote[1]{Here, the amplitude is set to 1 which is usually given to maximize the signal reflection strength.}, where $j$ is the imaginary unit, $\theta_n\in[0, 2\pi]$ is the phase shift of the $n$th element and thus $\phi_n$ has unit modulus, i.e., $|\phi_n|=1$. $\mathbf{\Phi}=\mathrm{diag}\{\phi_1, \phi_2, \cdots, \phi_N\}$ is the RED matrix. By appropriately tuning the phase shifts of the reflective elements of the IRS, the interference imposed on the PR can be mitigated whereas the useful signal received at the SR can be strengthened \cite{Lei2020Intelligent}. Denote the channels ST-IRS, ST-PR and ST-SR $k$ (SR $k$ is the $k$th SR, $k\in \mathcal{K}=\{1, 2, \cdots, K\}$) by $\mathbf{F}\in \mathbb{C}^{N\times M_t}$, $\mathbf{g}_d\in \mathbb{C}^{M_t\times 1}$ and $\mathbf{h}_{d,k}\in\mathbb{C}^{M_t\times 1}$, respectively. The reflective channels IRS-PR and IRS-SR $k$ are denoted by $\mathbf{g}_r\in\mathbb{C}^{N\times 1}$ and $\mathbf{h}_{r,k}\in\mathbb{C}^{N\times 1}$, respectively. Note that the channels $\mathbf{g}_d$ and $\mathbf{g}_r$ are related to the PR. The PR-related channels $\mathbf{g}_d$ and $\mathbf{g}_r$ are imperfectly estimated, since the CSI of the PR-related channels, including those from the ST to the IRS and from the IRS to the PRs, is usually more difficult to acquire than that of SR-related channels due to the lack of cooperation between licensed users and unlicensed users in CR systems \cite{Zhang2009RobustC}. In this paper, we assumed that the underlay spectrum access scheme is adopted. Based on that, practical energy detection is assumed to be used for spectrum sensing individually by SRs. To alleviate the hidden terminal problem, cooperative spectrum sensing can be adopted in which measurements of multiple spectrum sensors are combined to make a final decision on the existence of a primary signal.
\par The desired signal of SR $k$ is denoted by $s_k\in\mathbb{C}$ satisfying $\mathbb{E}[s_ks_k^*]=1$ and $\mathbb{E}[s_is_j^*]=0 (i\neq j)$, which has a corresponding TPC vector $\mathbf{w}_k\in\mathbb{C}^{M_t\times 1}$. The TPC matrix is denoted by $\mathbf{W}=[\mathbf{w}_1, \mathbf{w}_2, \cdots, \mathbf{w}_K]\in\mathbb{C}^{M_t\times K}$. Then, the transmit signal from the ST can be written as
$\mathbf{x}=\sum_{k=1}^K\mathbf{w}_ks_k$. The received signal at SR $k$ is
$y_k=(\mathbf{h}_{d,k}^\mathrm{H}+\mathbf{h}_{r,k}^\mathrm{H}\mathbf{\Phi}\mathbf{F})\mathbf{x}+n_{s_k}$, where $n_{s_k}\sim \mathcal{CN}(0,\sigma_{s_k}^2)$ is the equivalent noise which captures the joint effect of the thermal noise and the received interference from the primary transmitters \cite{Xu2020ResourceJ,Zhang2009RobustC}.
The received interference signal at the PR is given by
$y_p=(\mathbf{g}_{d}^\mathrm{H}+\mathbf{g}_{r}^\mathrm{H}\mathbf{\Phi}\mathbf{F})\mathbf{x}+n_{p}$, where $n_{p}\sim \mathcal{CN}(0,\sigma_{p}^2)$. Hence, the signal to interference plus noise ratio (SINR) of SR $k$ is
\begin{equation}\label{eq4}
\begin{split}
SINR_k&=\frac{|(\mathbf{h}_{d,k}^\mathrm{H}+\mathbf{h}_{r,k}^\mathrm{H}\mathbf{\Phi}\mathbf{F})\mathbf{w}_k|^2}{\|(\mathbf{h}_{d,k}^\mathrm{H}+\mathbf{h}_{r,k}
^\mathrm{H}\mathbf{\Phi}\mathbf{F})\mathbf{W}_{-k}\|_2^2+\sigma_{s_k}^2}\\
&=\frac{|(\mathbf{h}_{d,k}^\mathrm{H}
+\boldsymbol{\phi}^\mathrm{H}\mathrm{diag}(\mathbf{h}_{r,k}^\mathrm{H})\mathbf{F})\mathbf{w}_k|^2}{\|(\mathbf{h}_{d,k}^\mathrm{H}
+\boldsymbol{\phi}^\mathrm{H}\mathrm{diag}(\mathbf{h}_{r,k}
^\mathrm{H})\mathbf{F})\mathbf{W}_{-k}\|_2^2+\sigma_{s_k}^2},
\end{split}
\end{equation}
where $\mathbf{W}_{-k}=[\mathbf{w}_1, \cdots, \mathbf{w}_{k-1}, \mathbf{w}_{k+1}, \cdots, \mathbf{w}_K]$ and $\boldsymbol{\phi}=[\phi_1, \phi_2, \cdots, \phi_N]^\mathrm{T}$.
\par The interference temperature (IT) imposed on the PR from the ST is
\begin{equation}\label{eq_separate}
IT\!=\!\|(\mathbf{g}_{d}^\mathrm{H}+\mathbf{g}_{r}^\mathrm{H}\mathbf{\Phi}\mathbf{F})\mathbf{W}\|_2^2\!=\!\|(\mathbf{g}_{d}^\mathrm{H}
+\boldsymbol{\phi}^\mathrm{H}\mathbf{G}_{r})\!\mathbf{W}\!\|_2^2\!=\!\|\tilde{\boldsymbol{\phi}}^\mathrm{H}\!\mathbf{G}\!\mathbf{W}\!\|_2^2,
\end{equation}
where $\mathbf{G}_r=\mathrm{diag}(\mathbf{g}_r^\mathrm{H})\mathbf{F}\in \mathbb{C}^{N\times M_t}$ is regarded as the cascaded ST-IRS-PR channel, $\mathbf{G}=\left[\mathbf{G}_r^\mathrm{H}\,\,\mathbf{g}_d\right]^\mathrm{H}\in \mathbb{C}^{(N+1)\times M_t}$ is an equivalent combined channel integrating $\mathbf{g}_d$ and $\mathbf{G}_r$, and $\tilde{\boldsymbol{\phi}}=[\phi_1, \phi_2, \cdots, \phi_N, 1]^\mathrm{T}\in \mathbb{C}^{(N+1)\times 1}$. 
\subsection{Channel Uncertainty Models}
The channel uncertainty is caused by the imperfect estimation of the PR-related channels. If the direct channel $\mathbf{g}_d$ and the cascaded ST-IRS-PR channel $\mathbf{G}_r$ are separately estimated, the channels can be modeled as
\begin{equation}\label{eq_SDCun}
\mathbf{g}_d=\widehat{\mathbf{g}}_d+\triangle\mathbf{g}_d,\mathbf{G}_r=\widehat{\mathbf{G}}_r+\triangle\mathbf{G}_r,
\end{equation}
where $\widehat{\mathbf{g}}_d$ and $\widehat{\mathbf{G}}_r$ are estimated CSIs for the direct channel $\mathbf{g}_d$ and cascaded channel $\mathbf{G}_r$, respectively. $\triangle\mathbf{g}_d$ and $\triangle\mathbf{G}_r$ are corresponding CSI errors. If we substitute the channel estimations of (\ref{eq_SDCun}) into the equivalent combined channel, i.e., $\mathbf{G}$, then we have
\begin{equation}\label{eq_ch}
\mathbf{G}=\widehat{\mathbf{G}}+ \bigtriangleup \mathbf{G},
\end{equation}
where $\widehat{\mathbf{G}}=\left[\widehat{\mathbf{G}}_r^\mathrm{H}\,\,\,\widehat{\mathbf{g}}_d\right]^\mathrm{H}$ is regarded as the estimated combined CSI at the ST, $\triangle \mathbf{G}=\left[\triangle\mathbf{G}_r^\mathrm{H} \,\,\, \triangle\mathbf{g}_d \right]^\mathrm{H}$ is regarded as the combined CSI error matrix. In this paper, two different types of error models are investigated to describe the above CSI errors.
\subsubsection{Bounded CSI Error Model}
In this model, the CSI errors of the direct channel and cascaded channel are assumed to be bounded in the region as follows:
\begin{equation}\label{eq_gdgr}
\|\mathbf{\triangle g}_d\|_2\leq \epsilon_d, \|\mathbf{\triangle G}_r\|_\mathrm{F}\leq \epsilon_r,
\end{equation}
where $\epsilon_d$ and $\epsilon_r$ are the radii of bounded regions of CSI errors.
\subsubsection{Statistical CSI Error Model}
In this model, the CSI error vectors of $\triangle\mathbf{G}_r$ and $\triangle\mathbf{g}_d$ are assumed to follow the CSCG distributions with zero mean and covariance matrices of $\mathbf{\Sigma}_{g_r}$ and $\mathbf{\Sigma}_{g_d}$
, i.e.,
\begin{equation}\label{eq7}
\begin{split}
\mathbf{\triangle g}_d\sim\mathcal{CN}(\mathbf{0},\mathbf{\Sigma}_{g_d}),\;\mathrm{vec}(\mathbf{\triangle G}_r)\sim\mathcal{CN}(\mathbf{0},\mathbf{\Sigma}_{g_r}),
\end{split}
\end{equation}
where $\mathbf{\Sigma}_{g_d}\in \mathbb{C}^{M_t\times M_t}$ and $\mathbf{\Sigma}_{g_r}\in \mathbb{C}^{NM_t\times NM_t}$ are positive semidefinite matrices. $\mathbf{\triangle g}_d$ is independent of $\mathrm{vec}(\mathbf{\triangle G}_r)$.
According to the relationship of $\triangle \mathbf{G}=\left[\triangle\mathbf{G}_r^\mathrm{H} \,\,\, \triangle\mathbf{g}_d \right]^\mathrm{H}$, it can be verified that $\triangle\mathbf{G}$ also follows the CSCG distribution with zero mean and covariance matrix of $\mathbf{\Sigma}$, i.e.,
\begin{equation}\label{eqCSCG}
\mathrm{vec}(\mathbf{\triangle G})\sim\mathcal{CN}(\mathbf{0},\mathbf{\Sigma}),
\end{equation}
where $\mathbf{\Sigma}=\mathrm{diag}\{\mathbf{\Sigma}_{g_r},\mathbf{\Sigma}_{g_d}\}$. That means the model of (\ref{eq7}) is equivalent to that of (\ref{eqCSCG}). We will only focus on the latter one in this paper when we consider the statistical CSI error model.
\par The robust beamforming design will be first investigated under SCD scheme based on the bounded CSI error model shown in (\ref{eq_gdgr}). It is regarded as the worst-case beamforming to guarantee the SINR and the IT requirements for any channel realization. Then we will investigate the robust beamforming design under STA scheme based on the statistical CSI error model shown in (\ref{eqCSCG}). This type of robust beamforming is designed to guarantee the outage probability requirement of the PU's transmission.
\section{SCD Robust Design Based on Bounded Error Model}\label{section_SCD}
In this section, we investigate the SCD robust designing scheme based on the error model $\|\mathbf{\triangle g}_d\|_2\leq \epsilon_d$ and $\|\mathbf{\triangle G}_r\|_\mathrm{F}\leq \epsilon_r$. The original problem is firstly transformed into a deterministic problem by considering the worst-case IT constraint.
\subsection{Optimization Problem for SCD Scheme}
\par The problem with the aim of minimizing the ST's total transmit power by optimizing the TPC matrix $\mathbf{W}$ and the RED matrix $\mathbf{\Phi}$ can be formulated as
\begin{subequations}\label{eq_problem1}
\begin{align}
\mathcal{P}1\quad&\min\limits_{\mathbf{W},\mathbf{\Phi}}\quad \|\mathbf{W}\|_\mathrm{F}^2\label{eq_problem1_a}\\
\mathrm{s.t.}\quad &\frac{|(\mathbf{h}_{d,k}^\mathrm{H}
+\boldsymbol{\phi}^\mathrm{H}\mathrm{diag}(\mathbf{h}_{r,k}^\mathrm{H})\mathbf{F})\mathbf{w}_k|^2}{\|(\mathbf{h}_{d,k}^\mathrm{H}
+\boldsymbol{\phi}^\mathrm{H}\mathrm{diag}(\mathbf{h}_{r,k}
^\mathrm{H})\mathbf{F})\mathbf{W}_{-k}\|_2^2+\sigma_{s_k}^2}\geq \gamma_k,\forall k\in \mathcal{K},\label{eq_problem1_b}\\
&\|(\mathbf{g}_{d}^\mathrm{H}+\boldsymbol{\phi}^\mathrm{H}\mathbf{G}_{r})\mathbf{W}\|_2^2\leq \Gamma\!_{\mathrm{th}}, \|\mathbf{\triangle g}_d\|_2\leq \epsilon_d, \|\mathbf{\triangle G}_r\|_\mathrm{F}\leq \epsilon_r, \label{eq_problem1_c}\\
&|\phi_n|=1, \forall n\in \mathcal{N}. \label{eq_problem1_d}
\end{align}
\end{subequations}
In this problem, the error of the direct ST-PR channel and the error of the cascaded ST-BS-IRS channel are separately considered. The SINR constraint (\ref{eq_problem1_b}) is non-convex. Fortunately, we can use its successive convex approximation (SCA) version. Then, the remaining challenge is the bounded CSI error model of PU-related channels which makes the problem more complicated. In this section, we will employ Schur's complement and the General sign-definiteness principle to deal with the bounded CSI error.
\par By applying the Schur's complement lemma \cite{Boyd2004Convex}, the IT inequality constraint (\ref{eq_problem1_c}) can be rewritten as the following matrix inequality
\begin{equation}\label{IT_schur}
    \left[
      \begin{array}{cc}
        \Gamma\!_{\mathrm{th}} & \mathbf{b}^\mathrm{H} \\
        \mathbf{b} & \mathbf{I} \\
      \end{array}
    \right]\succeq \mathbf{0},
\end{equation}
where $\mathbf{b}=\left[(\mathbf{g}_d^\mathrm{H}+\boldsymbol{\phi}^\mathrm{H}\mathbf{G}_r)\mathbf{W}\right]^\mathrm{H}$.
By substituting $\mathbf{g}_d=\widehat{\mathbf{g}}_d+\triangle\mathbf{g}_d$ and $\mathbf{G}_r=\widehat{\mathbf{G}}_r+\triangle\mathbf{G}_r$ into (\ref{IT_schur}), we have
\begin{equation}\label{IT_schur_1}
\begin{split}\left[\!\!\begin{array}{cc}\Gamma\!_{\mathrm{th}}\!\! & \!\!\widehat{\mathbf{b}}^\mathrm{H} \\ \widehat{\mathbf{b}} \!\!&\!\! \mathbf{I} \\ \end{array}\!\!\right]
\!\!\succeq \!\!&-\left[\!\!\!\begin{array}{c}\mathbf{0} \\ \mathbf{W}^\mathrm{H} \\\end{array}\!\!\!\right]\!\!\triangle \mathbf{G}_r^\mathrm{H}[\begin{array}{cc}\!\!\boldsymbol{\phi} \!\!&\!\! \mathbf{0} \!\!\\\end{array}]
\!-\!\left[\!\!\!\begin{array}{c}\boldsymbol{\phi}^\mathrm{H} \\ \mathbf{0} \\\end{array}\!\!\!\right]\!\!\triangle\mathbf{G}_r[\!\!\begin{array}{cc}\mathbf{0}\!\! &\!\! \mathbf{W} \\
\end{array}\!\!]\\
&-\left[\!\!\begin{array}{c}\mathbf{0} \\ \mathbf{W}^\mathrm{H} \\ \end{array}
  \!\!\right][\!\!\begin{array}{cc}\triangle\mathbf{g}_d & \mathbf{0} \\
\end{array}\!\!]
\!-\!\left[\!\!\begin{array}{c}\triangle\mathbf{g}_d^\mathrm{H} \\
                     \mathbf{0}\\
                   \end{array}\!\!\right][\!\!\begin{array}{cc}\mathbf{0} & \mathbf{W} \\
                          \end{array}\!\!],
\end{split}
\end{equation}
where $\widehat{\mathbf{b}}=[(\widehat{\mathbf{g}}_d^\mathrm{H}+\boldsymbol{\phi}^\mathrm{H}\widehat{\mathbf{G}}_r)\mathbf{W}]^\mathrm{H}$.
\begin{lemma}(General sign-definiteness principle)\label{lemma0}
For a given set of matrices $\{\mathbf{Z}, \mathbf{U}_i, \mathbf{V}_i, i=1, \cdots, P\}$, where $\mathbf{Z}$ is Hermitian matrix, the following inequality
\begin{equation}\label{eq_lemma0}
\mathbf{Z}\succeq\sum_{i=1}^P(\mathbf{U}_i^\mathrm{H}\mathbf{X}_i\mathbf{V}_i+\mathbf{V}_i^\mathrm{H}\mathbf{X}_i^\mathrm{H}\mathbf{U}_i), \|\mathbf{X}_i\|\leq \epsilon_i, \forall i,
\end{equation}
holds if and only if there exist real values $\rho_i\geq 0, \forall i$ such that
\begin{equation}\label{lemma}
    \left[\!\!
      \begin{array}{cccc}
        \mathbf{Z}\!-\!\!\sum_{i=1}^P\rho_i\mathbf{V}_i^\mathrm{H}\mathbf{V}_i &-\epsilon_1\mathbf{U}_1^\mathrm{H} & \cdots & -\epsilon_P\mathbf{U}_P^\mathrm{H}\\
        -\epsilon_1\mathbf{U}_1 & \rho_1\mathbf{I} & \cdots& \mathbf{0} \\
        \vdots &  \vdots & \ddots & \vdots\\
        -\epsilon_P\mathbf{U}_P & \mathbf{0} & \cdots & \rho_P\mathbf{I} \\
      \end{array}
    \!\!\right]\!\!\succeq\!\mathbf{0}.
\end{equation}
\end{lemma}
\par \emph{Proof:} Please refer to \cite{Gharavol2013The}.$\hfill \blacksquare$
\par By comparing (\ref{eq_lemma0}) with (\ref{IT_schur_1}), we set the following equalities
\begin{equation*}
    \begin{split}
     & \mathbf{Z}\!=\! \left[\!\!\begin{array}{cc}\Gamma\!_{\mathrm{th}} & \widehat{\mathbf{b}}^\mathrm{H} \\ \widehat{\mathbf{b}} & \mathbf{I} \\ \end{array}\!\!\right], P\!=\!2,\epsilon_1\!=\!\epsilon_r,\epsilon_2\!=\!\epsilon_d,
        \mathbf{U}_1^\mathrm{H}\!=\!\mathbf{U}_2^\mathrm{H}\!=\!-\!\left[\!\!\begin{array}{c}\mathbf{0} \\ \mathbf{W}^\mathrm{H} \\ \end{array}\!\!\right],\\
        &\mathbf{X}_1=\triangle\mathbf{G}_r^\mathrm{H}, \mathbf{X}_2=\triangle\mathbf{g}_d,
        \mathbf{V}_1=[\begin{array}{cc}\boldsymbol{\phi} & \mathbf{0} \\ \end{array}],
        \mathbf{V}_2=[\begin{array}{cc}1 & \mathbf{0} \\ \end{array}].
    \end{split}
\end{equation*}
According to Lemma \ref{lemma0}, the equivalent form of the worst-case IT constraint (\ref{IT_schur_1}) is
\begin{equation}\label{IT_lemma2}
    \left[
      \begin{array}{cccc}
        \Gamma\!_{\mathrm{th}}-\rho_1N-\rho_2 & \widehat{\mathbf{b}}^\mathrm{H} & \mathbf{0}_{1\times M_t} & \mathbf{0}_{1\times M_t} \\
        \widehat{\mathbf{b}} & \mathbf{I}_K & \epsilon_r\mathbf{W}^\mathrm{H} & \epsilon_d\mathbf{W}^\mathrm{H} \\
        \mathbf{0}_{M_t\times 1} & \epsilon_r\mathbf{W} & \rho_1\mathbf{I}_{M_t} & \mathbf{0} \\
        \mathbf{0}_{M_t\times 1} & \epsilon_d\mathbf{W} & \mathbf{0} & \rho_2\mathbf{I}_{M_t} \\
      \end{array}
    \right]\succeq \mathbf{0}.
\end{equation}
Therefore, Problem $\mathcal{P}1$ can be reformulated as
\begin{equation}\label{eq_problem2}
\mathcal{P}1'\;\;\min\limits_{\mathbf{W},\mathbf{\Phi}, \boldsymbol{\rho}=\{\rho_1,\rho_2\}}\; \|\mathbf{W}\|_\mathrm{F}^2\;\;
\mathrm{s.t.}\; (\ref{eq_problem1_b}),(\ref{IT_lemma2}), (\ref{eq_problem1_d}),\boldsymbol{\rho}\!\geq\! 0.
\end{equation}
Since $\mathbf{W}$ and $\mathbf{\Phi}$ are coupled in Problem $\mathcal{P}1'$, the BCD method is used to alternately optimize $\mathbf{W}$ and $\mathbf{\Phi}$.
\subsection{Optimize $\mathbf{W}$ with Fixed $\mathbf{\Phi}$ for SCD Scheme}
For convenience, we firstly deal with SINR constraint (\ref{eq_problem1_b}) when $\mathbf{\Phi}$ is fixed. Then we apply the convex optimization approach to solve this subproblem.
\subsubsection{Deal with SINR constraint (\ref{eq_problem1_b})}
\par By defining $\widehat{\mathbf{h}}_{\mathbf{\Phi},k}^\mathrm{H}=\mathbf{h}_{d,k}^\mathrm{H}+\boldsymbol{\phi}^\mathrm{H}\mathrm{diag}(\mathbf{h}_{r,k}^\mathrm{H})\mathbf{F}$ and $\widehat{\mathbf{H}}_{\mathbf{\Phi},k}=\widehat{\mathbf{h}}_{\mathbf{\Phi},k}\widehat{\mathbf{h}}_{\mathbf{\Phi},k}^\mathrm{H}$, we have
\begin{equation}\label{eq8}
|(\mathbf{h}_{d,k}^\mathrm{H}+\boldsymbol{\phi}^\mathrm{H}\mathrm{diag}(\mathbf{h}_{r,k}^\mathrm{H})\mathbf{F})\mathbf{w}_k|^2
=\mathbf{w}_k^\mathrm{H}\widehat{\mathbf{H}}_{\mathbf{\Phi},k}\mathbf{w}_k,
\end{equation}
\begin{equation}\label{eq9}
\begin{split}
\left\|(\mathbf{h}_{d,k}^\mathrm{H}+\boldsymbol{\phi}^\mathrm{H}\mathrm{diag}(\mathbf{h}_{r,k}^\mathrm{H})\mathbf{F})\mathbf{W}_{-k}\right\|_2^2
=\sum_{j\neq k}^K\mathbf{w}_j^\mathrm{H}\widehat{\mathbf{H}}_{\mathbf{\Phi},k}\mathbf{w}_j.
\end{split}
\end{equation}
By substituting (\ref{eq8}) and (\ref{eq9}) into (\ref{eq_problem1_b}), we can rewrite the SINR constraint (\ref{eq_problem1_b}) as
\begin{equation}\label{eq10}
\frac{\mathbf{w}_k^\mathrm{H}\widehat{\mathbf{H}}_{\mathbf{\Phi},k}\mathbf{w}_k}{\sum_{j\neq k}^K\mathbf{w}_j^\mathrm{H}\widehat{\mathbf{H}}_{\mathbf{\Phi},k}\mathbf{w}_j+\sigma_{s_k}^2}\geq\gamma_k,\forall k\in \mathcal{K},
\end{equation}
which are non-convex. By using the first-order Taylor expansion, $\mathbf{w}_k^\mathrm{H}\widehat{\mathbf{H}}_{\mathbf{\Phi},k}\mathbf{w}_k$ can be lower bounded linearly by
$2\mathrm{Re}\{
\mathbf{w}_k^{(t)^\mathrm{H}}\widehat{\mathbf{H}}_{\mathbf{\Phi},k}\mathbf{w}_k\}-\mathbf{w}_k^{(t)^\mathrm{H}}\widehat{\mathbf{H}}_{\mathbf{\Phi},k}\mathbf{w}_k^{(t)}$,
where $\mathbf{w}_k^{(t)}$ is the optimal value of the TPC at the $t$th iteration. Then we can construct an SCA version of SINR constraint as
\begin{equation}\label{eq_gam}
2\mathrm{Re}(\mathbf{w}_k^{(t)^\mathrm{H}}\widehat{\mathbf{H}}_{\mathbf{\Phi},k}\mathbf{w}_k)-\gamma_k\sum_{j\neq k}^K\mathbf{w}_j^\mathrm{H}\widehat{\mathbf{H}}_{\mathbf{\Phi},k}\mathbf{w}_j\geq \widetilde{\gamma}_k,
\end{equation}
where $\widetilde{\gamma}_k=\mathbf{w}_k^{(t)^\mathrm{H}}\widehat{\mathbf{H}}_{\mathbf{\Phi},k}\mathbf{w}_k^{(t)}
+\gamma_k\sigma_{s_k}^2$.
\subsubsection{Subproblem to Optimize $\mathbf{W}$} When RED matrix is fixed, the SINR constraint (\ref{eq_problem1_b}) can be replaced by (\ref{eq_gam}), then the subproblem for optimizing $\mathbf{W}$ is given by
\begin{equation}\label{eq_problem2_1}
\mathcal{P}1.1\quad\min\limits_{\{\mathbf{w}_k\}, \boldsymbol\rho}\quad \sum_{k=1}^K\mathbf{w}_k^\mathrm{H}\mathbf{w}_k\quad
\mathrm{s.t.}\:\:(\ref{eq_gam}), (\ref{IT_lemma2}), \boldsymbol{\rho}\geq 0.
\end{equation}
IT constraint (\textrm{\ref{IT_lemma2}}) is a linear matrix inequality (LMI) when $\mathbf{\Phi}$ is fixed. This problem is an SOCP problem with respect to $\{\mathbf{w}_k, \forall k\}$, $\rho_1$ and $\rho_2$, which can be solved by using CVX.
\subsection{Optimize $\mathbf{\Phi}$ with Fixed $\mathbf{W}$ for SCD Scheme}
When TPC matrix $\mathbf{W}$ is given, the subproblem degenerates into a feasibility-check problem. We firstly transform the non-convex SINR constraint into a convex form. Then we reformulate the optimization problem to find the optimal $\mathbf{\Phi}$.
\subsubsection{Deal with SINR constraint (\ref{eq_problem1_b})}
\par Denoting $\mathbf{H}_{r,k}=\mathrm{diag}(\mathbf{h}_{r,k}^\mathrm{H})\mathbf{F}$, the constraint (\ref{eq_problem1_b}) can be rewritten as
\begin{equation}\label{eq_phi1}
\boldsymbol{\phi}^\mathrm{H}\mathbf{\Omega}_k\boldsymbol{\phi}\!-\!\gamma_k\boldsymbol{\phi}^\mathrm{H}\mathbf{\Omega}_{-k}\boldsymbol{\phi}
\!+\!2\mathrm{Re}\{\boldsymbol{\omega}_k^\mathrm{H}\boldsymbol{\phi}\}+\!\omega_k\!\geq\!\gamma_k\sigma_{s_k}^2, \forall k\!\in\!\mathcal{K},
\end{equation}
where
$\mathbf{\Omega}_k\!=\!\mathbf{H}_{r,k}\mathbf{w}_k\mathbf{w}_k^\mathrm{H}\mathbf{H}_{r,k}^\mathrm{H}, \omega_k=\mathbf{h}_{d,k}^\mathrm{H}\mathbf{w}_k\mathbf{w}_k^\mathrm{H}\mathbf{h}_{d,k}-\gamma_k\mathbf{h}_{d,k}^\mathrm{H}\sum_{j\neq k}^K(\mathbf{w}_j\mathbf{w}_j^\mathrm{H})\mathbf{h}_{d,k},
\mathbf{\Omega}_{-k}=\mathbf{H}_{r,k}\sum_{j\neq k}^K(\mathbf{w}_j\mathbf{w}_j^\mathrm{H})\mathbf{H}_{r,k}^\mathrm{H}$, and $\boldsymbol{\omega}_k=\mathbf{H}_{r,k}\mathbf{w}_k\mathbf{w}_k^\mathrm{H}\mathbf{h}_{d,k}-\gamma_k\mathbf{H}_{r,k}\sum_{j\neq k}^K(\mathbf{w}_j\mathbf{w}_j^\mathrm{H})\mathbf{h}_{d,k}$. As a constraint, (\ref{eq_phi1}) is still non-convex.
\par By using the first-order Taylor inequality,
$\boldsymbol{\phi}^\mathrm{H}\mathbf{\Omega}_k\boldsymbol{\phi}$ can be lower bounded linearly by $2\mathrm{Re}\{
\boldsymbol{\phi}^{(t)^\mathrm{H}}\mathbf{\Omega}_k\boldsymbol{\phi}\} -\boldsymbol{\phi}^{(t)^\mathrm{H}}\mathbf{\Omega}_k\boldsymbol{\phi}^{(t)}$.
We can construct an SCA version of SINR constraint (\ref{eq_phi1}) as
\begin{equation}\label{eq_phi2_2}
\gamma_k\boldsymbol{\phi}^\mathrm{H}\mathbf{\Omega}_{-k}\boldsymbol{\phi}-2\mathrm{Re}\{(\boldsymbol{\omega}_k^\mathrm{H}
+\boldsymbol{\phi}^{(t)^\mathrm{H}}\mathbf{\Omega}_k)\boldsymbol{\phi}\}\leq\overline{\gamma}_k,
\end{equation}
where $\boldsymbol{\phi}^{(t)}$ is solution of the $t$th iteration and $\overline{\gamma}_k=\omega_k-\boldsymbol{\phi}^{(t)^\mathrm{H}}\mathbf{\Omega}_k\boldsymbol{\phi}^{(t)}-\gamma_k\sigma_{s_k}^2$.
\subsubsection{Subproblem to Optimize $\mathbf{\Phi}$}
When $\mathbf{W}$ is fixed, the SINR constraint can be approximated by (\ref{eq_phi2_2}). To deal with this feasibility-check problem, the slack variables $\boldsymbol{\varphi}=[\varphi_1, \varphi_2, \cdots, \varphi_K]$ are introduced to tighten the constraints (\ref{eq_phi1}). Then the quadratic inequality (\ref{eq_phi2_2}) is modified as
\begin{equation}\label{eq_phi2_3}
\gamma_k\boldsymbol{\phi}^\mathrm{H}\mathbf{\Omega}_{-k}\boldsymbol{\phi}-2\mathrm{Re}\{(\boldsymbol{\omega}_k^\mathrm{H}
+\boldsymbol{\phi}^{(t)^\mathrm{H}}\mathbf{\Omega}_k)\boldsymbol{\phi}\}\leq\overline{\gamma}_k-\varphi_k, \forall k\in\mathcal{K}.
\end{equation}
\par Therefore, for the feasibility-check problem, both the SINR constraint given by (\ref{eq_phi2_3}) and the IT constraint given by (\ref{IT_lemma2}) are convex. Then the subproblem can be reformulated as
\begin{equation}\label{eq_problem2_2}
\mathcal{P}1.2\quad \max\limits_{\boldsymbol{\phi}, \boldsymbol{\varphi}}\quad \sum\limits_{k=1}^{K} \varphi_k \quad
\mathrm{s.t.}\:\:(\ref{eq_phi2_3}),(\ref{IT_lemma2}),(\ref{eq_problem1_d}),\boldsymbol{\varphi}\geq 0.
\end{equation}
The only non-convexity of Problem $\mathcal{P}1.2$ is from the unit-modulus constraint of $\boldsymbol{\phi}$ in (\ref{eq_problem1_d}). Then we adopt the penalty CCP method to deal with it \cite{Zhou2020Framework}. According to the penalty CCP principle, the non-convex constraint (\ref{eq_problem1_d}) can be first equivalently transformed into $1\!\leq\!|\phi_n|^2\!\leq\!1$. The non-convex part can be linearized by $|\phi_n^{(t)}|^2\!-\!2\mathrm{Re}(\phi_n^\ast\phi_n^{(t)})\!\leq\!-1$. Then, we can reformulate Problem $\mathcal{P}1.2$ as
\begin{subequations}\label{eq_problem1_2_1}
	\begin{align}
	\mathcal{P}1.2\_1:\quad &\max\limits_{\boldsymbol{\phi}, \boldsymbol{\varphi}, \boldsymbol{\tau}}\quad  \sum\limits_{k=1}^K \varphi_k-\kappa\sum\limits_{n=1}^{2N} \tau_n\label{eq_problem1_2_1a}\\
	\mathrm{s.t.}\quad&(\ref{eq_phi2_3}), (\ref{IT_lemma2}), \boldsymbol{\varphi}\geq 0, \boldsymbol{\tau}\geq 0,\label{eq_problem1_2_1a+}\\
	&|\phi_n^{(t)}|^2-2\mathrm{Re}(\phi_n^\ast\phi_n^{(t)})\leq \tau_n-1,\forall n\in \mathcal{N},\label{eq_problem1_2_1c}\\
	&|\phi_n|^2\leq1+\tau_{N+n},\forall n\in \mathcal{N},\label{eq_problem1_2_1d}
	\end{align}
\end{subequations}
where $\boldsymbol{\tau}\!=\![\tau_1,\!\cdots\!,\tau_{2N}\!]^\mathrm{T}$ are slack variables. $\kappa$ is a penalty multiplier to scale the penalty item $\sum_{n=1}^{2N} \!\tau_n$ which can control the feasibility of $\boldsymbol{\phi}$ combining with adjustable $\kappa$. This subproblem is an SOCP problem which can be solved by using CVX.
\par
\begin{figure}[!t]
	\centering
	\includegraphics[width=2.5in]{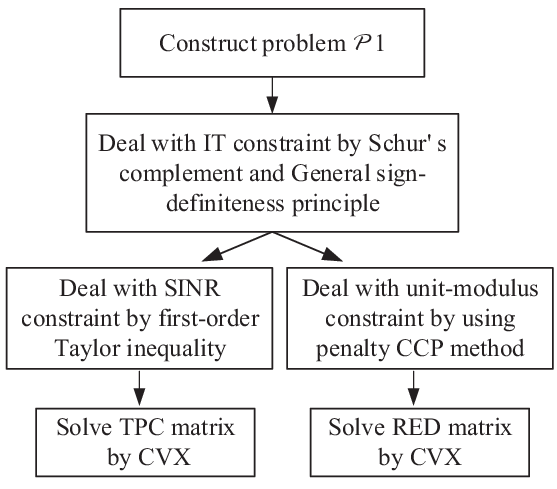}
	\caption{The key steps and methodologies involved in SCD scheme.}
	\label{fig-flowchart1}
\end{figure}

 For the convenience of understanding, a flow chart is provided in Fig. \ref{fig-flowchart1} to summarize the key steps of the whole optimization procedure for SCD scheme.
\subsection{Overall Algorithm for SCD Scheme}
\subsubsection{Convergence analysis}
\begin{algorithm}[htb]
	\caption{Optimizing $\mathbf{W}$ and $\mathbf{\Phi}$ for SCD}\label{alg:1}
	\begin{algorithmic}[1]
		\STATE
		Initialize $\mathbf{w}_k^{(0)}$ and $\boldsymbol{\phi}^{(0)}$ in feasible region for $\forall k\in\mathcal{K}$, set maximum iteration number $t_{max}$, target convergence accuracy $\zeta$, $l_\kappa>1$, $\kappa_{max}$ and $t=0$.
		\STATE \textbf{Repeat}
		\STATE Update $\mathbf{w}_k^{(\!t+1\!)}$ by solving Problem $\mathcal{P}1.1$; $n\!=\!0$,  $\boldsymbol{\phi}^{(\!n\!)}\!=\!\boldsymbol{\phi}^{(\!t\!)}$;
		\STATE \quad\textbf{Repeat}
		\STATE \quad Update $\boldsymbol{\phi}^{(n+1)}$ by solving Problem $\mathcal{P}1.2\_1$;
		\STATE \quad$\kappa^{(n+1)}=\max\{l_\kappa\kappa^{(n)}, \kappa_{max}\}$;
		\STATE \quad\textbf{Until} $\|\boldsymbol{\tau}\|_1\leq l_1$ and $\|\boldsymbol{\phi}^{(n+1)}-\boldsymbol{\phi}^{(n)}\|_1\leq l_2$;
		\STATE $\boldsymbol{\phi}^{(t+1)}=\boldsymbol{\phi}^{(n+1)}$;
		\STATE Calculate $\|\mathbf{W}^{(t+1)}\|_\mathrm{F}^2$ from (\ref{eq_problem1_a}) in Problem $\mathcal{P}1$;
		\STATE \textbf{Until} $t>t_{max}$ or $\frac{\|\mathbf{W}^{(t+1)}\|_\mathrm{F}^2-\|\mathbf{W}^{(t)}\|_\mathrm{F}^2}{\|\mathbf{W}^{(t+1)}\|_\mathrm{F}^2}<\zeta$.
	\end{algorithmic}
\end{algorithm}
\par The overall algorithm for the SCD scheme is provided in Algorithm \ref{alg:1}.
The initialization value of $\mathbf{w}_k^{(0)}$ and $\boldsymbol{\phi}^{(0)}$ can be obtained from the optimal solution of feasibility checking problem which will be discussed in Section \ref{section6}.
In the iteration for updating $\kappa$, we set a sufficiently small value $l_1$ to check whether $\|\boldsymbol\tau\|_1<l_1$, which is one condition to stop the iteration. $\|\boldsymbol\tau\|_1<l_1$ satisfies the constraint (\ref{eq_problem1_d}) in (\ref{eq_problem2_2}). Another iteration stopping condition is $\|\boldsymbol{\phi}^{(n+1)}-\boldsymbol{\phi}^{(n)}\|_1\leq l_2$, where $l_2$ is a small value. It can be verified that the sequences of the objective function value produced by Algorithm \ref{alg:1} are guaranteed to converge. For the first subproblem optimizing $\mathbf{W}$, we show that the solution sequence $\{\mathbf{W}^{(t+1)}, t=1,2,\cdots \}$ is feasible and the objective function value sequence $\{f(\mathbf{W}^{(t)})=\|\mathbf{W}^{(t)}\|_\mathrm{F}^2, t=1,2,\cdots\}$ is monotonically decreasing. Define the functions $g(\mathbf{w}_k)\triangleq \mathbf{w}_k^\mathrm{H}\widehat{\mathbf{H}}_{\mathbf{\Phi},k}\mathbf{w}_k$ and $\overline{g}(\mathbf{w}_k|\mathbf{w}_k^{(t)})\triangleq -\mathbf{w}_k^{(t)^\mathrm{H}}\widehat{\mathbf{H}}_{\mathbf{\Phi},k}\mathbf{w}_k^{(t)}+2\mathrm{Re}\{
\mathbf{w}_k^{(t)^\mathrm{H}}\widehat{\mathbf{H}}_{\mathbf{\Phi},k}\mathbf{w}_k\}$. Assume that $\mathbf{W}^{(t+1)}$ is a feasible solution of Problem $\mathcal{P}1.1$, and thus $\mathbf{W}^{(t+1)}$ satisfies the constraints (\textrm{\ref{eq_gam}}) and (\textrm{\ref{IT_lemma2}}). By substituting $\mathbf{W}^{(t+1)}$ into (\textrm{\ref{eq_gam}}), we have $\frac{\overline{g}(\mathbf{w}_k^{(t+1)}|\mathbf{w}_k^{(t)})}{\sum_{j\neq k}^K\mathbf{w}_j^{(t+1)^\mathrm{H}}\widehat{\mathbf{H}}_{\mathbf{\Phi},k}\mathbf{w}_j^{(t+1)}+\sigma_{s_k}^2}\geq\gamma_k,\forall k$. According to the first-order Taylor inequality, we have $\overline{g}(\mathbf{w}_k^{(t+1)}|\mathbf{w}_k^{(t)})\leq g(\mathbf{w}_k^{(t+1)})$. Thus, $\frac{g(\mathbf{w}_k^{(t+1)})}{\sum_{j\neq k}^K\mathbf{w}_j^{(t+1)^\mathrm{H}}\widehat{\mathbf{H}}_{\mathbf{\Phi},k}\mathbf{w}_j^{(t+1)}+\sigma_{s_k}^2}\geq\gamma_k$ holds, which means that the sequence $\{\mathbf{W}^{(t+1)}, t=1,2,\cdots \}$ satisfies the constraint (\ref{eq10}) that is the SINR constraint (\ref{eq_problem1_b}). As $\mathbf{W}^{(t+1)}$ is the globally optimal solution of Problem $\mathcal{P}1.1$, $f(\mathbf{W}^{(t+1)})\leq f(\mathbf{W}^{(t)})$ holds. Thus, the objective function value sequence $\{f(\mathbf{W}^{(t)}), t=1,2,\cdots\}$ is guaranteed to converge. For the second subproblem of optimizing $\mathbf{\Phi}$, the convergence can be easily verified by using the same analysis approach.
\subsubsection{Complexity analysis}
Since the proposed Algorithm \ref{alg:1} involves SOC, LMI and linear constraints which can be solved by a standard interior point method, the general expression of the computational complexity of which is given by
\begin{equation}\label{complexity0}
    \mathcal{O}((\sum_{J}J_jb_j+2I)^{0.5}n(n^2+n\sum_{J}J_jb_j^2+\sum_{J}J_jb_j^3+n\sum_{I}I_ia_i^2)),
\end{equation}
where we ignore the complexity of linear constraints, $n$ is the number of variables, $I_i$ is the number of SOC constraints with the size of $a_i$ and $\sum I_i=I$, $J_j$ is the number of LMI constraints with size of $b_j$ and $\sum J_j=J$. Based on the above general expression, the complexity for solving Problem $\mathcal{P}1.1$ is $o^{scd}_\mathbf{w}=\mathcal{O}((2M_t+3K+1)^{0.5}M_tK(M_tK^2+M_tK(2M_t+K+1)^2+(2M_t+K+1)^3+K^2M_t^3))$, the complexity of Problem $\mathcal{P}1.2$ is $o^{scd}_\mathbf{\Phi}=\mathcal{O}(\log(1/{\min\{l_1,l_2\}})(2N+3K+1)^{0.5}N(N(K+1)^2+N(K+1)^3+KN^3+2N^2))$. Thus, the overall complexity for solving Problem $\mathcal{P}1$ is $o^{scd}_\mathbf{w}+o^{scd}_\mathbf{\Phi}$.
\section{STA Robust Design Based on Statistical Error Model}\label{STA}
In this section, we optimize the robust beamforming based on statistical CSI error model where the uncertain CSI $\triangle\mathbf{G}$ satisfies the distribution as shown in (\ref{eqCSCG}), i.e., $\mathrm{vec}(\mathbf{\triangle G})\sim\mathcal{CN}(\mathbf{0},\mathbf{\Sigma})$.
\subsection{Optimization Problem for STA Scheme}
\par The power minimization problem considering statistical CSI error can be formulated as
\begin{subequations}\label{eq_problem2}
\begin{align}
\mathcal{P}2:\quad&\min\limits_{\mathbf{W},\mathbf{\Phi}}\quad \|\mathbf{W}\|_\mathrm{F}^2\label{eq_problem2_a}\\
\mathrm{s.t.}\quad &\frac{|(\mathbf{h}_{d,k}^\mathrm{H}
+\boldsymbol{\phi}^\mathrm{H}\mathrm{diag}(\mathbf{h}_{r,k}^\mathrm{H})\mathbf{F})\mathbf{w}_k|^2}{\|(\mathbf{h}_{d,k}^\mathrm{H}
+\boldsymbol{\phi}^\mathrm{H}\mathrm{diag}(\mathbf{h}_{r,k}
^\mathrm{H})\mathbf{F})\mathbf{W}_{-k}\|_2^2+\sigma_{s_k}^2}\geq \gamma_k,\forall k\in \mathcal{K},\label{eq_problem2_b}\\
&\mathrm{Pr}\left\{\left\|(\tilde{\boldsymbol{\phi}}^\mathrm{H}\widehat{\mathbf{G}}+\tilde{\boldsymbol{\phi}}^\mathrm{H}\triangle\mathbf{G})\mathbf{W}\right\|_2^2\leq \Gamma\!_{\mathrm{th}}\right\}\geq 1-\beta,\nonumber\\
&\mathrm{vec}(\mathbf{\triangle G})\sim\mathcal{CN}(\mathbf{0},\mathbf{\Sigma}),\label{eq_problem2_c}\\
&|\phi_n|=1, \forall n\in \mathcal{N}, \label{eq_problem2_d}
\end{align}
\end{subequations}
where $\tilde{\boldsymbol{\phi}}$ is defined as in (\ref{eq_separate}), $0\leq\beta\leq 1$ is the outage probability when the IT imposed on the PR exceeds the threshold $\Gamma\!_{\mathrm{th}}$. The real challenge to solve this optimization problem is the statistical CSI error model of (\ref{eq_problem2_c}) since the SINR constraint (\ref{eq_problem2_b}) and the unit-modulus constraint (\ref{eq_problem2_d}) of the phase shifts at the IRS can be handled in the same way as in Problem $\mathcal{P}1$. Hence, in this section, we will first reformulate the probability constraint by using Chi-square distribution method. The related result is shown in Proposition \ref{proposition3}.
\par Firstly, we reformulate the IT inequality constraint. According to the triangle inequality
$\|(\tilde{\boldsymbol{\phi}}^\mathrm{H}\widehat{\mathbf{G}}+\tilde{\boldsymbol{\phi}}^\mathrm{H}\triangle\mathbf{G})\mathbf{W}\|_2^2\leq \|\tilde{\boldsymbol{\phi}}^\mathrm{H}\widehat{\mathbf{G}}\mathbf{W}\|_2^2+\|\tilde{\boldsymbol{\phi}}^\mathrm{H}\triangle\mathbf{G}
\mathbf{W}\|_2^2+2\|\tilde{\boldsymbol{\phi}}^\mathrm{H}\widehat{\mathbf{G}}\mathbf{W}\|_2\|\tilde{\boldsymbol{\phi}}^\mathrm{H}\triangle\mathbf{G}
\mathbf{W}\|_2\leq 2\left( \|\tilde{\boldsymbol{\phi}}^\mathrm{H}\widehat{\mathbf{G}}\mathbf{W}\|_2^2+\|\tilde{\boldsymbol{\phi}}^\mathrm{H}\triangle\mathbf{G}
\mathbf{W}\|_2^2\right)$, the following probabilistic relationship holds
\begin{equation}\label{eq_2_3}
\begin{split}
\mathrm{Pr}&\left\{\|(\tilde{\boldsymbol{\phi}}^\mathrm{H}\widehat{\mathbf{G}}+\tilde{\boldsymbol{\phi}}^\mathrm{H}\triangle\mathbf{G})\mathbf{W}\|_2^2\leq \Gamma_{\mathrm{th}}\right\}\geq \\ &\mathrm{Pr}\left\{\|\tilde{\boldsymbol{\phi}}^\mathrm{H}\widehat{\mathbf{G}}\mathbf{W}\|_2^2+\|\tilde{\boldsymbol{\phi}}^\mathrm{H}\triangle\mathbf{G}
\mathbf{W}\|_2^2\leq \Gamma\right\}.
\end{split}
\end{equation}
where $\Gamma=\frac{\Gamma_{\mathrm{th}}}{2}$. 
Thus, the IT inequality constraint can be approximated by
\begin{equation}\label{eq_2_4}
\mathrm{Pr}\left\{\|\tilde{\boldsymbol{\phi}}^\mathrm{H}\triangle\mathbf{G}
\mathbf{W}\|_2^2\leq\Gamma-\|\tilde{\boldsymbol{\phi}}^\mathrm{H}\widehat{\mathbf{G}}\mathbf{W}\|_2^2\right\}\geq 1-\beta.
\end{equation}
Based on the above analysis, the following proposition holds.
\begin{prop}\label{proposition3}
Assume that $\mathrm{vec}(\triangle\mathbf{G})\in\mathbb{C}^{(N+1)M_t\times 1}$ is a complex Gaussian vector satisfying $\mathrm{vec}(\mathbf{\triangle G})\sim\mathcal{CN}(\mathbf{0},\mathbf{\Sigma})$ as in (\ref{eqCSCG}). The sufficient condition for the probabilistic constraint (\ref{eq_2_4}) to hold is that
\begin{equation}\label{eq_2_5}
\vartheta F_{2(N+1)M_t}^{-1}(1-\beta)\|\mathbf{W}\|_\mathrm{F}^2+\|\tilde{\boldsymbol{\phi}}^\mathrm{H}\widehat{\mathbf{G}}\mathbf{W}\|_2^2\leq \Gamma,
\end{equation}
\end{prop}
where $\vartheta=(N+1)\lambda_{max}(\mathbf{\Sigma})$, $\lambda_{max}(\cdot)$ is the maximum eigenvalue and $F_{n}^{-1}(\cdot)$ is the inverse Chi-square cumulative distribution function with $n$ degrees of freedom.
\par\emph{Proof:} Please see Appendix \ref{append4}. $\hfill \blacksquare$
\par Note that when $\beta\!=\!0$, the constraint (\ref{eq_problem2_c}) means the worst-case for the statistical error model. If $\beta\!=\!1$, (\ref{eq_problem2_c}) can be removed because the probability constraint is always satisfied. In this case, the PR cannot be protected.
\par By replacing the IT constraint (\ref{eq_problem2_c}) with (\ref{eq_2_5}), Problem $\mathcal{P}2$ is reformulated as
\begin{equation}\label{eq_problem3}
\mathcal{P}2':\quad\min\limits_{\mathbf{W},\mathbf{\Phi}}\quad \|\mathbf{W}\|_\mathrm{F}^2\quad
\mathrm{s.t.}\quad (\ref{eq_problem2_b}), (\ref{eq_2_5}), (\ref{eq_problem2_d}).
\end{equation}
This problem is also intractable because $\mathbf{W}$ and $\mathbf{\Phi}$ are coupled in (\ref{eq_problem2_b}) and (\ref{eq_2_5}). The BCD algorithm is adopted to optimize these two variables alternately.
\subsection{Optimize $\mathbf{W}$ with Fixed $\mathbf{\Phi}$ for STA Scheme}
\par When $\mathbf{\Phi}$ is fixed, the IT constraint (\ref{eq_2_5}) can be rewritten as
\begin{equation}\label{eq_2_6}
\sum_{k=1}^K\mathbf{w}_k^\mathrm{H}(\vartheta F_{2(N+1)M_t}^{-1}(1-\beta)\mathbf{I}+\widetilde{\mathbf{G}}_{\tilde{{\boldsymbol{\phi}}}})\mathbf{w}_k\leq \Gamma.
\end{equation}
where $\widetilde{\mathbf{G}}_{\tilde{{\boldsymbol{\phi}}}}=
\widehat{\mathbf{G}}^\mathrm{H}\tilde{\boldsymbol{\phi}}\tilde{\boldsymbol{\phi}}^\mathrm{H}\widehat{\mathbf{G}}$.
The inequality constraint (\ref{eq10}) in SCD scheme can be directly used to replace the SINR constraint (\ref{eq_problem2_b}).
Then the problem for optimizing $\mathbf{W}$ is reformulated as
\begin{equation}\label{eq_problem3_1}
\mathcal{P}2.1:\quad\min\limits_{\{\mathbf{w}_k\}}\quad \sum_{k=1}^K\mathbf{w}_k^\mathrm{H}\mathbf{w}_k\quad
\mathrm{s.t.}\:\: (\ref{eq10}), (\ref{eq_2_6}).
\end{equation}
The SINR constraint (\ref{eq10}) is non-convex. We can use SDR approach to solve this problem. Specifically, define $\mathbf{S}_k=\mathbf{w}_k\mathbf{w}_k^\mathrm{H},\forall k\in \mathcal{K}$ with the constraints that $\mathrm{rank}(\mathbf{S}_k)=1,\forall k \in \mathcal{K}$. Then, Problem $\mathcal{P}2.1$ can be reformulated as
\begin{subequations}\label{eq_problem3_1_1}
	\begin{align}
	 \mathcal{P}2.1\_1:&\quad\min\limits_{\{\mathbf{S}_k\succeq\mathbf{0}\}_{k\in\mathcal{K}}}\quad \sum_{k=1}^K\mathrm{tr}(\mathbf{S}_k)\label{eq_problem3_1_a}\\
	\mathrm{s.t.}\quad
	\mathrm{tr}&(\mathbf{S}_k\widehat{\mathbf{H}}_{\mathbf{\Phi},k}\!)\!\geq\! \gamma_k\!\left(\sum_{j\neq k}^K\!\mathrm{tr}(\mathbf{S}_j\widehat{\mathbf{H}}_{\mathbf{\Phi},k})\!+\!\sigma_{s_k}^2\! \right) ,\forall k,\label{eq_problem3_1_b}\\
	\sum_{k=1}^K&\mathrm{tr}\left(\mathbf{S}_k\left(\vartheta F_{2(N+1)M_t}^{-1}(1-\beta)\mathbf{I}+\widetilde{\mathbf{G}}_{\tilde{{\boldsymbol{\phi}}}}\right)\right)\leq \Gamma, \label{eq_problem3_1_c}\\
	&\mathrm{rank}(\mathbf{S}_k)=1, \forall k \in\mathcal{K}.\label{eq_problem3_1_d}
	\end{align}
\end{subequations}
However, Problem $\mathcal{P}2.1\_1$ is still non-convex due to the rank-one constraint. We further relax this rank-one constraint and obtain the following optimization problem:
\begin{equation}\label{eq_problem3_1_2}
\mathcal{P}2.1\_2:\quad\min\limits_{\{\mathbf{S}_k\succeq \mathbf{0}\}_{k\in\mathcal{K}}}\quad \sum_{k=1}^K\mathrm{tr}(\mathbf{S}_k)\quad
\mathrm{s.t.}\:\: (\ref{eq_problem3_1_b}), (\ref{eq_problem3_1_c}).
\end{equation}
Obviously, Problem $\mathcal{P}2.1\_2$ is an SDP problem, which is convex and can be effectively solved by using the standard tools such as CVX.
\par In general, the optimal solution obtained from SDP problem may not satisfy the rank-one constraints. Fortunately, for Problem $\mathcal{P}2.1\_2$, we can prove that the relaxation of the non-convex
constraints is tight. Denote the optimal solution of $\mathcal{P}2.1\_2$ as $\mathbf{S}_k^{\star}, \forall k\in \mathcal{K}$, we have the following
theorem.
\begin{theorem}\label{theorem1}
	The optimal solution obtained from the SDP Problem $\mathcal{P}2.1\_2$ is guaranteed to satisfy	the rank-one constraints, i.e., $\mathrm{rank}(\mathbf{S}_k^\star)=1, \forall k\in \mathcal{K}$.
\end{theorem}
\par
\emph{Proof:} Please see Appendix \ref{append3}.$\hfill \blacksquare$
\par Since the rank of $\mathbf{S}_k, \forall k\in\mathcal{K}$ is equal to one, we can use the simple singular value decomposition operation to obtain the optimal beam-vector of Problem $\mathcal{P}2.1$.
\subsection{Optimize $\mathbf{\Phi}$ with Fixed $\mathbf{W}$ for STA Scheme}
\par When $\mathbf{W}$ is fixed, this subproblem degenerates into a feasibility-check problem. The SINR constraint (\ref{eq_phi2_2}) can be directly used to replace (\ref{eq_problem2_b}). The IT constraint (\ref{eq_2_5}) can be rewritten as
\begin{equation}\label{eq_2_7}
\sum_{k=1}^K\!\tilde{\boldsymbol{\phi}}^\mathrm{H}\widehat{\mathbf{G}}\mathbf{w}_k\!\mathbf{w}_k^\mathrm{H}\widehat{\mathbf{G}}^\mathrm{H}\tilde{\boldsymbol{\phi}}\!\leq\! \Gamma\!-\!\vartheta F_{2(N+1)M_t}^{-1}(1\!-\!\beta)\left\|\mathbf{W}\right\|_\mathrm{F}^2.
\end{equation}
Let
$\mathbf{X}\triangleq\sum_{k=1}^K\widehat{\mathbf{G}}\mathbf{w}_k\mathbf{w}_k^\mathrm{H}\widehat{\mathbf{G}}^\mathrm{H}$ which is a Hermitian matrix. Extract the first $N$ rows and $N$ columns elements of $\mathbf{X}$ as a sub-matrix $\mathbf{B}$. Denote the vector consisting of elements in the ($N+1$)th column from the first row to the $N$th row of $\mathbf{X}$ by $\mathbf{b}$. Denote the vector consisting of elements in the ($N+1$)th row from the first column to the $N$th column of $\mathbf{X}$ by $\mathbf{c}$ and denote the ($N+1$)th row and ($N+1$)th column element by $b_{N+1}$. Thus, we have
\begin{equation}\label{eq_phi4}
\sum_{k=1}^K\tilde{\boldsymbol{\phi}}^\mathrm{H}\widehat{\mathbf{G}}\mathbf{w}_k\mathbf{w}_k^\mathrm{H}\widehat{\mathbf{G}}^\mathrm{H}\tilde{\boldsymbol{\phi}}=\boldsymbol{\phi}^\mathrm{H}\mathbf{B}\boldsymbol{\phi}+2\mathrm{Re}\{\mathbf{b}^\mathrm{H}\boldsymbol{\phi}\}+b_{N+1}.
\end{equation}
Finally, the IT constraint (\ref{eq_2_7}) is reformulated as
\begin{equation}\label{eq_2_8}
\boldsymbol{\phi}^\mathrm{H}\mathbf{B}\boldsymbol{\phi}+2\mathrm{Re}\{\mathbf{b}^\mathrm{H}\boldsymbol{\phi}\}\!\leq\! \Gamma\!-\!\vartheta F_{2(N+1)M_t}^{-1}\!(1\!-\!\beta)\!\left\|\mathbf{W}\right\|_\mathrm{F}^2\!-\!b_{N\!+\!1}.
\end{equation}
\par By introducing the slack variables to the SINR constraint as in Problem $\mathcal{P}1.2$, the subproblem is
\begin{equation}\label{eq_problem3_2}
\mathcal{P}2.2:\max\limits_{\boldsymbol{\phi}, \boldsymbol{\varphi}}\quad \sum\limits_{k=1}^{K} \varphi_k\quad
\mathrm{s.t.}\quad(\ref{eq_phi2_3}),(\ref{eq_2_8}),(\ref{eq_problem2_d}),\boldsymbol{\varphi}\geq 0.
\end{equation}
Problem $\mathcal{P}2.2$ can be solved by using the same method with $\mathcal{P}1.2$, here we omit it for simplicity.
\par
\begin{figure}[!t]
		\centering
		\includegraphics[width=2.5in]{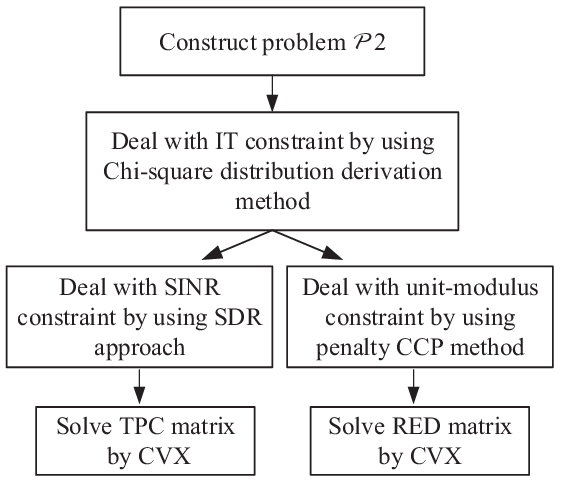}
		\caption{The key steps and methodologies involved in STA scheme.}
		\label{fig-flowchart2}
	\end{figure}

 For the convenience of understanding, a flow chart is provided in Fig. \ref{fig-flowchart2} to summarize the key steps of the whole optimization procedure for STA scheme.
\subsection{Overall Algorithm for STA Scheme}
\begin{algorithm}[htb]
\caption{Optimizing $\mathbf{W}$ and $\mathbf{\Phi}$ for STA}\label{alg:3}
\begin{algorithmic}[1]
\STATE
Initialize $\mathbf{w}_k^{(0)}$ and  $\boldsymbol{\phi}^{(0)}$ in feasible region for $\forall k\in\mathcal{K}$, set maximum iteration number $t_{max}$, target convergence accuracy $\zeta$ and $t=0$.
\STATE \textbf{Repeat}
\STATE Calculate $\mathbf{w}_k^{(t+1)}$ by solving Problem $\mathcal{P}2.1$;
\STATE Calculate $\boldsymbol{\phi}^{(t+1)}$ by solving Problem $\mathcal{P}2.2$;
\STATE Calculate $\|\mathbf{W}^{(t+1)}\|_\mathrm{F}^2$ from (\ref{eq_problem2_a}) in Problem $\mathcal{P}2$;
\STATE \textbf{Until} $t>t_{max}$ or $\frac{\|\mathbf{W}^{(t+1)}\|_\mathrm{F}^2-\|\mathbf{W}^{(t)}\|_\mathrm{F}^2}{\|\mathbf{W}^{(t+1)}\|_\mathrm{F}^2}<\zeta$.
\end{algorithmic}
\end{algorithm}
The overall algorithm for solving Problem $\mathcal{P}2$ is given in Algorithm \ref{alg:3}.
The initialization value of $\mathbf{w}_k^{(0)}$ and $\boldsymbol{\phi}^{(0)}$ can be obtained from the optimal solution of feasibility checking problem which will be discussed in Section \ref{section6}.
Since the proposed Algorithm \ref{alg:3} involving SOC, LMI and linear constraints can be addressed by a standard interior point method, the general expression of the computational complexity of which can be given as the equation (\ref{complexity0}). Based on that, the complexity of Problem $\mathcal{P}2.1$ is $o^{sta}_\mathbf{w}=\mathcal{O}((K+1)^{0.5}(K(K+1)^2M_t^{4.5}+K^3M_t^{3.5}))+\mathcal{O}(M_t^3)$ where the second term $\mathcal{O}(M_t^3)$ is produced by singular value decomposition operation of matrix solution. The complexity of Problem $\mathcal{P}2.2$ is $o^{sta}_\mathbf{\Phi}=\mathcal{O}(\log(1/{\min\{l_1,l_2\}})\sqrt{2}(K+N+1)^{0.5}[(K+1)N^4+2N^3])$. Then the overall complexity of Problem $\mathcal{P}2$ is $o^{sta}_\mathbf{w}+o^{sta}_\mathbf{\Phi}$. It can be also readily verified that the sequences of the objective function value produced by Algorithm \ref{alg:3} are guaranteed to converge.
\section{Feasibility Checking For Problems $\mathcal{P}_1$ and $\mathcal{P}_2$}\label{section6}
Due to the conflicting SINR and IT constraints, problems $\mathcal{P}1$ and $\mathcal{P}2$ may be infeasible for SCD scheme and STA scheme, respectively. Hence, we have to first check whether problems $\mathcal{P}1$ and $\mathcal{P}2$ are feasible or not. To this end, we construct the following two feasibility checking problems.
\subsection{Feasibility Checking for Problem $\mathcal{P}1$}	
By minimizing the interference imposed on the PU, we can construct the feasibility checking problem for SCD scheme as
\begin{subequations}\label{eq_prochk0}
	\begin{align}
	\overline{\mathcal{P}1}:&\quad\min\limits_{\mathbf{W},\mathbf{\Phi},\alpha}\quad \alpha\\
	\mathrm{s.t.}\;
	 &\|(\mathbf{g}_{d}^\mathrm{H}+\boldsymbol{\phi}^\mathrm{H}\mathbf{G}_{r})\mathbf{W}\|_2^2\leq \alpha\Gamma, \|\mathbf{\triangle g}_d\|_2\leq \epsilon_d, \|\mathbf{\triangle G}_r\|_\mathrm{F}\leq \epsilon_r, \label{eq_prochk0_c}\\
	&(\ref{eq_problem1_b}), (\ref{eq_problem1_d}),
	\end{align}
\end{subequations}
where $\alpha$ is an additional variable introduced to act as the indicators of the feasibility of
Problem $\mathcal{P}1$. If the optimal solution of $\alpha$ is smaller or equal to one, we claim that the main problem is feasible. Otherwise, the main Problem $\mathcal{P}1$ is declared to be infeasible.  By applying Schur's complement and Lemma \ref{lemma0}, the IT constraint (\ref{eq_prochk0_c}) of Problem $\overline{\mathcal{P}1}$ can be transformed into the following LMI
\begin{equation}\label{eq_prochk0_1}
\left[
\begin{array}{cccc}
\alpha\Gamma-\rho_1N-\rho_2 & \widehat{\mathbf{b}}^\mathrm{H} & \mathbf{0}_{1\times M_t} & \mathbf{0}_{1\times M_t} \\
\widehat{\mathbf{b}} & \mathbf{I}_K & \epsilon_r\mathbf{W}^\mathrm{H} & \epsilon_d\mathbf{W}^\mathrm{H} \\
\mathbf{0}_{M_t\times 1} & \epsilon_r\mathbf{W} & \rho_1\mathbf{I}_{M_t} & \mathbf{0} \\
\mathbf{0}_{M_t\times 1} & \epsilon_d\mathbf{W} & \mathbf{0} & \rho_2\mathbf{I}_{M_t} \\
\end{array}
\right]\succeq \mathbf{0}.
\end{equation}
Problem $\overline{\mathcal{P}1}$ can be reformulated as
\begin{equation}\label{eq_prochk1}
	 \overline{\mathcal{P}1'}:\quad\min\limits_{\mathbf{W},\mathbf{\Phi},\boldsymbol{\rho},\alpha}\quad \alpha \quad
	\mathrm{s.t.}\quad (\ref{eq_problem1_b}), (\ref{eq_prochk0_1}), (\ref{eq_problem1_d}), \boldsymbol{\rho}\geq 0.
\end{equation}
Since $\mathbf{W}$ and $\boldsymbol{\Phi}$ are coupled, the BCD approach is used again to alternately solve this problem. Based on this
idea, we present an iterative algorithm with two main steps.
\par In the first step, the TPC matrix $\mathbf{W}$ is optimized with fixed RED matrix $\boldsymbol{\Phi}$. Then
the alternative problem becomes
\begin{equation}\label{eq_prochk1.1}
	\overline{\mathcal{P}1}.1:\quad\min\limits_{\mathbf{W},\boldsymbol{\rho},\alpha} \quad\alpha \quad
	\mathrm{s.t.}\quad (\ref{eq_gam}), (\ref{eq_prochk0_1}), \boldsymbol{\rho}\geq 0.
\end{equation}
Problem $\overline{\mathcal{P}1}.1$ is an SOCP problem which is convex and can be effectively solved by using CVX. The optimal solution $\alpha$ obtained from this problem can be used to check the feasibility.
\par In the second step, the RED matrix is optimized when the TPC matrix, $\boldsymbol{\rho}$ and $\alpha$ are given. Then the alternative problem becomes
\begin{equation}\label{eq_prochk1.2}
\overline{\mathcal{P}1}.2:\quad\max\limits_{\boldsymbol{\Phi}, \boldsymbol{\varphi}}\quad \sum\limits_{k=1}^{K} \varphi_k\quad
	\mathrm{s.t.}\quad
(\ref{eq_phi2_3}), (\ref{eq_prochk0_1}), (\ref{eq_problem1_d}), \boldsymbol{\varphi}\geq 0,
\end{equation}
where $\boldsymbol{\varphi}=[\varphi_1, \varphi_2, \cdots, \varphi_K]$ are slack variables introduced into the SINR constraint. This problem is non-convex due to the unit-modulus constraint (\ref{eq_problem1_d}). Then we adopt the penalty CCP method to deal with this non-convex constraint. Problem $\overline{\mathcal{P}1}.2$ is transformed into an SOCP problem. The problem transformation process is similar to Problem $\mathcal{P}1.2$. Here we omit it for simplicity.
\par The feasibility checking algorithm for the main Problem $\mathcal{P}1$ is formally
described in Algorithm \ref{alg:4}.
\begin{algorithm}[htb]
	\caption{Feasibility Checking Algorithm for Problem $\mathcal{P}1$}\label{alg:4}
	\begin{algorithmic}[1]
		\STATE
		Initialize $\mathbf{W}^{(0)}$, $\boldsymbol{\Phi}^{(0)}$, $t_{max}$ and set $t=1$.
		\STATE Given RED matrix $\boldsymbol{\Phi}^{(t-1)}$, update TPC matrix by solving Problem $\overline{\mathcal{P}1}.1$. Denote the solution by $\alpha^\star$, $\mathbf{W}^\star$. Update $\mathbf{W}^{(t)}=\mathbf{W}^\star$, $\alpha^{(t)}=\alpha^\star$.
		\STATE If $\alpha^{(t)}\leq1$, declare that the main Problem $\mathcal{P}1$ is feasible and output these feasible $\mathbf{W}^{(t)}$ for the initialization of the main Problem $\mathcal{P}1$, and terminate. If $\alpha^{(t)}\geq1$ and $t\geq t_{max}$, declare that the main Problem $\mathcal{P}1$ is infeasible and terminate.
		Otherwise, continue.
		\STATE Given TPC $\mathbf{W}^{(t)}$, update RED matrix by solving Problem $\overline{\mathcal{P}1}.2$. Denote the solution by $\boldsymbol{\Phi}^\star$. Update $\boldsymbol{\Phi}^{(t)}\!=\!\boldsymbol{\Phi}^\star$.
		\STATE Let $t\leftarrow t+1$ and go to step 2.
	\end{algorithmic}
\end{algorithm}
\subsection{Feasibility Checking for Problem $\mathcal{P}2$}
If we set an outage probability $\beta$, the objective function can be regarded as the interference imposed on the PU which causes communication interruption of PU with probability $\beta$ where the uncertain CSI $\triangle\mathbf{G}$ satisfies the distribution of $\mathrm{vec}(\mathbf{\triangle G})\sim\mathcal{CN}(\mathbf{0},\mathbf{\Sigma})$. According to Proposition \ref{proposition3}, the objective function can be given by $\vartheta F_{2(N+1)M_t}^{-1}(1-\beta)\|\mathbf{W}\|_\mathrm{F}^2+\|\tilde{\boldsymbol{\phi}}^\mathrm{H}\widehat{\mathbf{G}}\mathbf{W}\|_2^2$. Hence, the feasibility checking problem for Problem $\mathcal{P}2$ is
\begin{subequations}\label{eq_prochk2}
	\begin{align}
	\overline{\mathcal{P}2}:\;&\min\limits_{\mathbf{W},\mathbf{\Phi}}\;\vartheta F_{2(N+1)M_t}^{-1}(1-\beta)\|\mathbf{W}\|_\mathrm{F}^2+\|\tilde{\boldsymbol{\phi}}^\mathrm{H}\widehat{\mathbf{G}}\mathbf{W}\|_2^2\label{eq_prochk2_a}\\
	\mathrm{s.t.}&\quad(\ref{eq_problem2_b}),  (\ref{eq_problem2_d}),
	\end{align}
\end{subequations}
where (\ref{eq_problem2_b}) is SINR constraint and (\ref{eq_problem2_d}) is unit-modulus constraint of phase shift at the IRS. The feasibility can be checked by judging the optimal objective value. If the optimal objective value is lower than IT threshold $\Gamma$, Problem $\overline{\mathcal{P}2}$ is feasible. Otherwise, it is infeasible. Since $\mathbf{W}$ and $\boldsymbol{\Phi}$ are coupled, we have to alternately solve this problem. We present the iterative algorithm with two main steps.
\par In the first step, TPC matrix is optimized with fixed RED matrix.
When $\boldsymbol{\Phi}$ is fixed, the objective function (\ref{eq_prochk2_a}) can be rewritten as
$\sum_{k=1}^K\mathbf{w}_k^\mathrm{H}(\vartheta F_{2(N+1)M_t}^{-1}(1-\beta)\mathbf{I}+\widetilde{\mathbf{G}}_{\tilde{{\boldsymbol{\phi}}}})\mathbf{w}_k$. The SINR constraints can be replaced by (\ref{eq10}). Then we can give the TPC matrix optimization problem by
\begin{equation}\label{eq_prochk2.1}
	\overline{\mathcal{P}2}.1:\;\min\limits_{\mathbf{W}}\; \sum_{k=1}^K\mathbf{w}_k^\mathrm{H}(\vartheta F_{2(N+1)M_t}^{-1}(1-\beta)\mathbf{I}+\widetilde{\mathbf{G}}_{\tilde{{\boldsymbol{\phi}}}})\mathbf{w}_k \quad
	\mathrm{s.t.}\; (\ref{eq10}).
\end{equation}
This problem can be solved by using the same SDR approach with Problem $\mathcal{P}2.1$. After defining $\mathbf{S}_k=\mathbf{w}_k\mathbf{w}_k^\mathrm{H}$ with the constraints that $\mathrm{rank}(\mathbf{S}_k)=1,\forall k \in \mathcal{K}$, Problem $\overline{\mathcal{P}2}.1$ can be transformed into
\begin{subequations}\label{eq_prochk2_1_1}
	\begin{align}
	\overline{\mathcal{P}2}.1\_1:\;&\min\limits_{\{\mathbf{S}_k\succeq\mathbf{0}\}}\! \sum_{k=1}^K\mathrm{tr}\!\left(\mathbf{S}_k\!\left(\vartheta F_{2(N+1)M_t}^{-1}(1\!-\!\beta)\mathbf{I}\!+\!\widetilde{\mathbf{G}}_{\tilde{{\boldsymbol{\phi}}}}\right)\!\right)\\
	\mathrm{s.t.}\;\mathrm{tr}&(\mathbf{S}_k\widehat{\mathbf{H}}_{\mathbf{\Phi},k})\!\geq\! \gamma_k\!\left(\!\sum_{j\neq k}^K\!\mathrm{tr}(\mathbf{S}_j\widehat{\mathbf{H}}_{\mathbf{\Phi},k})\!+\!\sigma_{s_k}^2\! \right)\!,\forall k,\label{eq_prochk2_1_b}\\
	&\mathrm{rank}(\mathbf{S}_k)=1, \forall k \in\mathcal{K}.\label{eq_prochk_1_d}
	\end{align}
\end{subequations}
We further relax this rank-one constraint and obtain an SDP optimization problem, which can be effectively solved by CVX. The solution of the SDP problem can be guaranteed to satisfy the rank-one constraints.
\par In the second step, RED matrix is optimized with fixed TPC matrix.
When $\mathbf{W}$ is given, the RED matrix optimization problem is given by
\begin{subequations}\label{eq_prochk2.2}
	\begin{align}
\overline{\mathcal{P}2}.2:\;&\min\limits_{\boldsymbol{\phi}}\boldsymbol{\phi}^\mathrm{H}\mathbf{B}\boldsymbol{\phi}\!+\!2\mathrm{Re}\{\mathbf{b}^\mathrm{H}\boldsymbol{\phi}\}\!+
\!\vartheta F_{2(N+1)M_t}^{-1}\!(1\!-\!\beta)\!\left\|\mathbf{W}\right\|_\mathrm{F}^2\\
	\mathrm{s.t.}\:
	 \gamma_k&\boldsymbol{\phi}^\mathrm{H}\mathbf{\Omega}_{-k}\boldsymbol{\phi}\!-\!2\mathrm{Re}\{\!(\boldsymbol{\omega}_k^\mathrm{H}
	 \!+\!\boldsymbol{\phi}^{(t)^\mathrm{H}}\!\mathbf{\Omega}_k)\boldsymbol{\phi}\}\!\leq\!\overline{\gamma}_k\!-\!\varphi_k, \forall k\!\in\!\mathcal{K},\label{eq_prochk2.2b}\\
	&|\phi_n|=1, \forall n\in \mathcal{N},\label{eq_prochk2.2c}
	\end{align}
\end{subequations}
where $\mathbf{B}$ and $\boldsymbol{\phi}$ in objective function is obtained from (\ref{eq_2_8}). This optimization problem can be solved by using the same method with Problem $\overline{\mathcal{P}1}.2$ where the penalty CCP method is used to deal with unit-modulus constraint. The problem can be transformed into an SOCP problem solved by CVX.
\par The feasibility checking algorithm for the main Problem $\mathcal{P}2$ is formally
described in Algorithm \ref{alg:5}.
\begin{algorithm}[htb]
	\caption{Feasibility Checking Algorithm for Problem $\mathcal{P}2$}\label{alg:5}
	\begin{algorithmic}[1]
		\STATE
		Initialize $\mathbf{W}^{(0)}$, $\boldsymbol{\Phi}^{(0)}$, $t_{max}$ and set $t=1$.
		\STATE Given RED matrix $\boldsymbol{\Phi}^{(t-1)}$, update TPC matrix by solving Problem $\overline{\mathcal{P}2}.1$. Denote the solution by $\mathbf{W}^\star$. Update $\mathbf{W}^{(t)}=\mathbf{W}^\star$.
		\STATE Given TPC matrix $\mathbf{W}^{(t)}$, update RED matrix by solving Problem $\overline{\mathcal{P}2}.2$. Denote the solution by $\boldsymbol{\Phi}^\star$. Update $\boldsymbol{\Phi}^{(t)}=\boldsymbol{\Phi}^\star$.
		Calculate the optimal value of objective function for Problem $\overline{\mathcal{P}2}$ and denote it as $f^\star$.
		\STATE If $f^\star\leq \Gamma$, declare the main Problem $\mathcal{P}2$ is feasible and output these feasible $\mathbf{W}^{(t)}$ and $\boldsymbol{\Phi}^{(t)}$ for the initialization of the main Problem $\mathcal{P}2$, and terminate. If $f^\star\geq \Gamma$ and $t\geq t_{max}$, declare the main Problem $\mathcal{P}2$ is infeasible and terminate.
		Otherwise, continue.
		\STATE Let $t\leftarrow t+1$ and go to step 2.
	\end{algorithmic}
\end{algorithm}
\section{Extension to Multiple PRs Scenario}\label{section5-}
In this section, we extend one PR scenario to multiple PRs scenario. Assume that $L$ PRs randomly locate in a cell region in the IRS-aided CR system. The other assumptions remain the same as in Fig. \ref{fig1}. The system parameters correspond to the $l$th PR ($l\!\in\! \mathcal{L}\!=\!\{1, 2, \!\cdots\!,\! L\}$) are direct channel $\mathbf{g}_{d,l}$, reflect channel $\mathbf{g}_{r,l}$ and IT threshold requirement $\Gamma_l$. By using $\mathbf{g}_{d,l}$, $\mathbf{g}_{r,l}$ and $\Gamma_l$ in IT expressions and channel uncertainty model, we can rewrite the optimization problem for the multiple PRs scenario.
\par Firstly, the IT imposed on PR $l$ from the ST is given by
\begin{equation}\label{ITl}
IT_l=\|\widetilde{\boldsymbol{\phi}}^{\mathrm{H}}\mathbf{G}_l\mathbf{W}\|_2^2, \forall l\in\mathcal{L},
\end{equation}
where $\mathbf{G}_l=[\mathbf{G}_{r,l}^{\mathrm{H}}\;\; \mathbf{g}_{d,l}]^{\mathrm{H}}\in\mathbb{C}^{(N+1)\times M_t} $ is an equivalent combined channel. $\mathbf{G}_{r,l}=\mathrm{diag}(\mathbf{g}_{r,l}^\mathrm{H})\mathbf{F}\in\mathbb{C}^{N\times M_t}$ is regarded as the cascaded ST-IRS-PR $l$ channel.
\par Secondly, for the bounded CSI error model, the radii of bounded regions of CSI errors for direct channel $\mathbf{g}_{d,l}$ and cascaded channel $\mathbf{G}_{r,l}$ can be denoted by $\epsilon_{d,l}$ and $\epsilon_{r,l}$, respectively, for $\forall l$. Then, the radius of the bounded region of CSI error for the combined channel $\mathbf{G}_l$ is denoted by $\epsilon_l$. For the statistical CSI error model, the distribution assumption in (\ref{eq7}) is rewritten as
\begin{equation}\label{eq7plus}
\begin{split}
\mathbf{\triangle g}_{d,l}\sim\mathcal{CN}(\mathbf{0},\mathbf{\Sigma}_{g_{d,l}})\;\mathrm{vec}(\mathbf{\triangle G}_{r,l})\sim\mathcal{CN}(\mathbf{0},\mathbf{\Sigma}_{g_{r,l}}),
\end{split}
\end{equation}
where $\mathbf{\Sigma}_{g_{d,l}}$ and $\mathbf{\Sigma}_{g_{r,l}}$ are covariance matrices for $\forall l\in\mathcal{L}$.
Thus, the CSCG distribution for $\triangle\mathbf{G}_l$ is given by
\begin{equation}\label{eqCSCGplus}
\mathrm{vec}(\mathbf{\triangle G}_l)\sim\mathcal{CN}(\mathbf{0},\mathbf{\Sigma}_l),\forall l\in\mathcal{L},
\end{equation}
where $\mathbf{\Sigma}_l=\mathrm{diag}\{\mathbf{\Sigma}_{g_{r,l}},\mathbf{\Sigma}_{g_{d,l}}\}$.
\par Based on the above assumptions, for the optimization problems of SCD and STA schemes, only the IT constraints need to be modified into multi-PRs form. Hence, constraint (\ref{eq_problem1_c}) in SCD optimization problem is rewritten as
\begin{equation}
\|(\mathbf{g}_{d,l}^\mathrm{H}+\boldsymbol{\phi}^\mathrm{H}\mathbf{G}_{r,l})\mathbf{W}\|_2^2\leq \Gamma_l, \|\mathbf{\triangle g}_{d,l}\|_2\leq \epsilon_{d,l}, \|\mathbf{\triangle G}_{r,l}\|_\mathrm{F}\leq \epsilon_{r,l}.
\end{equation}
Constraint (\ref{eq_problem2_c}) in STA optimization problem is rewritten as
\begin{equation}
\begin{split}
\mathrm{Pr}&\left\{\left\|(\tilde{\boldsymbol{\phi}}^\mathrm{H}\widehat{\mathbf{G}}_l+\tilde{\boldsymbol{\phi}}^\mathrm{H}\triangle\mathbf{G}_l)\mathbf{W}\right\|_2^2\leq \Gamma_l\right\}\geq 1-\beta_l,\\
&\mathrm{vec}(\mathbf{\triangle G}_l)\sim\mathcal{CN}(\mathbf{0},\mathbf{\Sigma}_l), \forall l \in \mathcal{L},
\end{split}
\end{equation}
where $\beta_l$ is outage probability due to secondary interference for PR $l$. The methods and mathematical derivations in single-PR optimization problems are applicable in multi-PRs optimization problems. Here, we omit it for simplicity. In simulation results, we will provide the impacts of number of PRs on system performance.
\section{Simulation Results}\label{section5--}
In this section, we firstly provide simulation results of one PR scenario to study the impacts of parameters on the robust beamforming design in the IRS-aided MISO CR system. Then we expand the simulation to multiple PRs scenario.
\begin{figure}[!t]
\vspace{-0.3cm}
\centering
\includegraphics[width=3.5in]{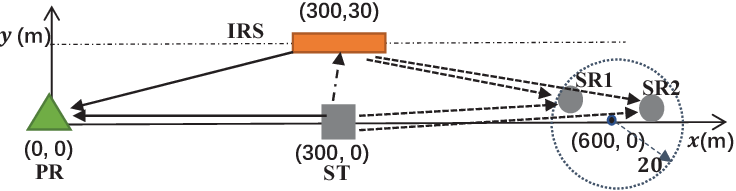}
\vspace{-0.5cm}
\caption{The simulated scenario of IRS-aided CR.}
\label{fig2_simu}
\end{figure}
The simulation scenario is shown in Fig. \ref{fig2_simu} where the PR, ST and IRS are located at the same horizontal line with locations given by (0 m, 0 m), (300 m, 0) and (300 m, 30 m), respectively. There are $K\!=\!2$ SRs randomly distributed in the region of a cell. The channel models are assumed to include large-scale fading and small-scale fading. Denote the large-scale path loss in dB by $P_\mathrm{L}\!\!=\!\!P_{\mathrm{L}_0}\!\!-\!10\alpha \log_{10}(\frac{d}{d_0})$, where $\alpha$ is the path loss exponent and $d$ is the transmission distance. The small-scale fading follows Rayleigh distribution. The minimum data rate requirement for SU is assumed to be the same at $r_k\!=\!r, \forall k\in\mathcal{K}$. The main parameters are collected in Table \ref{tab2}.
\begin{table}\centering
	\caption{Simulation parameters.}
	\label{tab2}
	\begin{tabular}{|c|c|}
		\hline
		Parameter & Value \\
		\hline
		PR location & (0 m, 0 m) \\
		IRS location & (300 m, 30 m) \\
		ST location & (300 m, 0 m) \\
		SR's cell location & (600 m, 0 m) \\
		Cell radius  & 20 m\\
		 Pass loss & $P_{\mathrm{L}_0}=30$ dB \\
		 Reference distance & $d_0=1$ m\\
	 Path loss exponents of IRS-related links & $\alpha_{\mathrm{IRS}}=2.2$\\
		Path loss exponent of the ST-PR link or ST-SR link & $\alpha_{\mathrm{STU}}=3.75$\cite{Xu2020ResourceJ,Zhou2020Framework} \\
		Outage probability &$\beta=0.05$\\
		Noise power density &  $-174$ dBm/Hz\\
		\hline
	\end{tabular}
\end{table}
\par The CSI error vectors $\Delta\mathbf{g}_d$ and $\mathrm{vec}(\Delta\mathbf{G}_r)$ are assumed to be correlated for the STA scheme. The covariance matrices of $\mathbf{\Sigma}_{g_r}$ and $\mathbf{\Sigma}_{g_d}$ are defined as
${\bf{\Sigma}}_{{g_r}}^{1/2}\!=\!{\sigma_{{g_r}}}\!{{\bf{C}}_{{g_r}}}$ and ${\bf{\Sigma }}_{{g_d}}^{1/2}\!=\!{\sigma_{{g_d}}}\!{{\bf{C}}_{{g_d}}}$. ${{\bf{C}}_{g_r}}\!\in\!\mathbb{C}^{N{M_t}\!\times \!N{M_t}}$ and ${\bf{C}}_{g_d}\!\in\!{\mathbb{C}}^{M_t\!\times\!M_t}$ are the correlation matrices, the elements of which are given by ${\left[ {{{\bf{C}}_{{g_r}}}} \right]_{m,n}}$, ${\left[ {{{\bf{C}}_{{g_d}}}} \right]_{m,n}}\!=\!c_\eta^{|m-n|}$ and ${c_\eta }$ is set as 0.9. To make the relative amount of correlated CSI errors comparable with the case of uncorrelated CSI errors, we assume that $\sigma^2_{g_r}\!=\!\delta^2_{g_r}\left\|\rm{vec}(\widehat{\mathbf{G}}_r)\right\|_2^2\left[\frac{\left\|\mathbf{I}_{NM_t}\right\|_\mathrm{F}}{\left\|\mathbf{C}^2_{g_r}\right\|_{\mathrm{F}}}\right]$  and $\sigma^2_{g_d}\!=\!\delta^2_{g_d}\left\|(\widehat{\mathbf{g}}_d)\right\|_2^2\left[\frac{\left\|\mathbf{I}_{M_t}\right\|_\mathrm{F}}{\left\|\mathbf{C}^2_{g_d}\right\|_{\mathrm{F}}}\right]$. $\delta_{g_d}\!\in\! [0, 1)$ and $\delta_{g_r}\!\in\! [0, 1)$ represent the channel uncertainty levels which measure the relative amount of CSI uncertainties, here we set $\delta_{g_d}\!=\!\delta_{g_r}\!=\!\delta_g$. The radii of the uncertainty regions are set as $\epsilon_d\!=\!\sqrt{\frac{\sigma_{g_d}^2}{2}F_{2M_t}^{-1}(1\!-\!\beta)}$ and $\epsilon_r\!=\!\sqrt{\frac{\sigma_{g_r}^2}{2}F_{2NM_t}^{-1}(1\!-\!\beta)}$, respectively. According to \cite{WangK2014Outage}, this bounded CSI error model provides a fair comparison between the performance of the worst-case robust design and the outage constrained robust design.

\subsection{The Convergence and Complexity of Algorithms}
\begin{figure}[htbp]
	\centering
	\begin{minipage}[t]{0.48\textwidth}
	\centering
\includegraphics[width=3.1in]{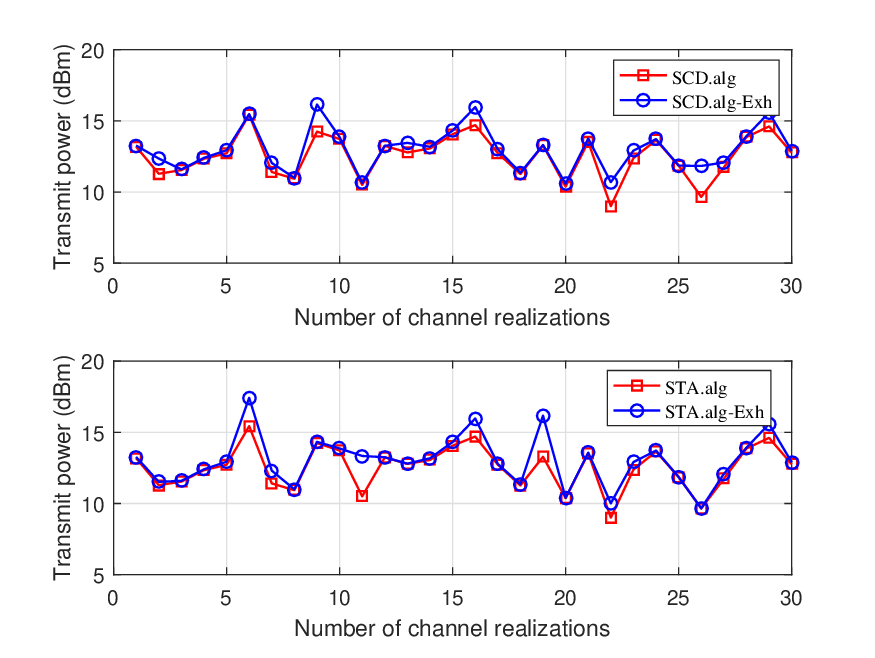}
\caption{The performance comparison of different initializations, with $r=0.5$, $M_t=4$, $\Gamma=-60$ dBm, $\beta=0.05$ and $\delta_{g}=0.05$.}
\label{optimility_channelreal}
	\end{minipage}
	\begin{minipage}[t]{0.48\textwidth}
	\centering
\includegraphics[width=3.1in]{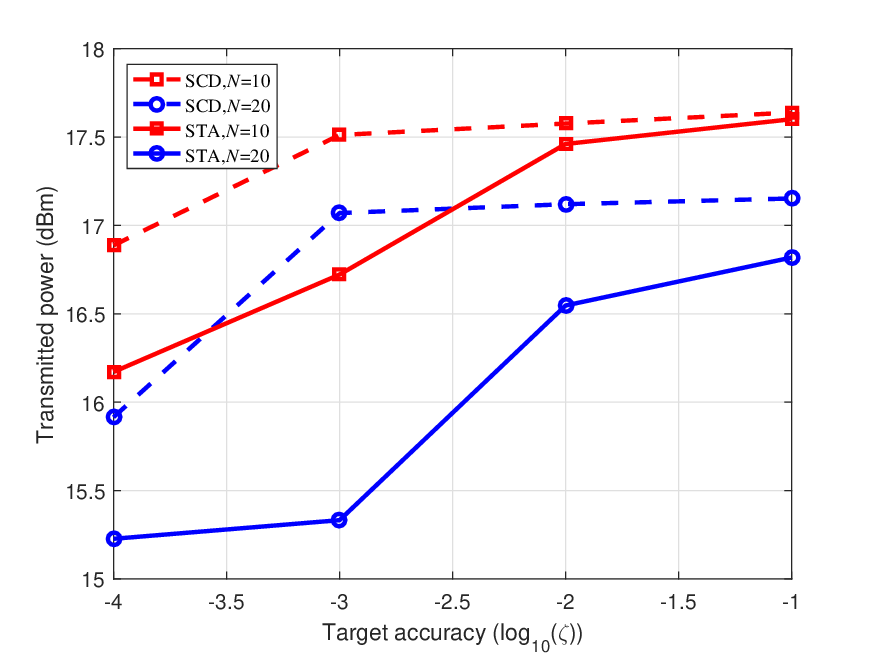}
\caption{Transmit power versus target accuracy, with $r=1$, $M_t=4$, $\Gamma=-90$ dBm, $\beta=0.05$ and $\delta_{g}=0.05$.}
\label{convergence_accuracy}
	\end{minipage}
\end{figure}
	
\begin{figure}[htbp]
\centering
\begin{minipage}[t]{0.48\textwidth}
	\centering
\includegraphics[width=3.1in]{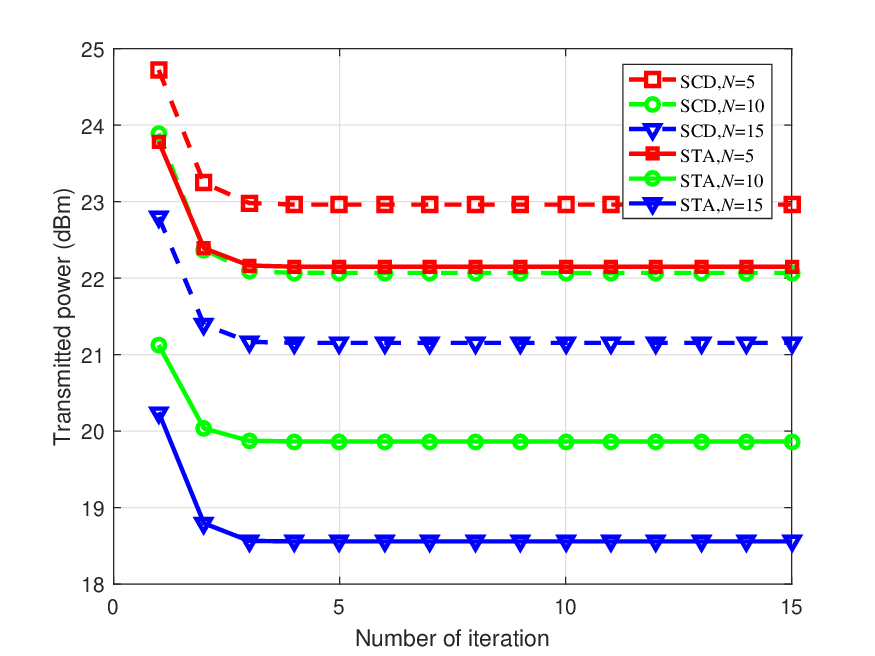}
\caption{Transmit power versus the number of iterations, with $r=2$, $M_t=4$, $\Gamma=-90$ dBm, $\beta=0.05$ and $\delta_{g}=0.05$.}
\label{convergence_iteration}
\end{minipage}
\begin{minipage}[t]{0.48\textwidth}
\centering
\includegraphics[width=3.1in]{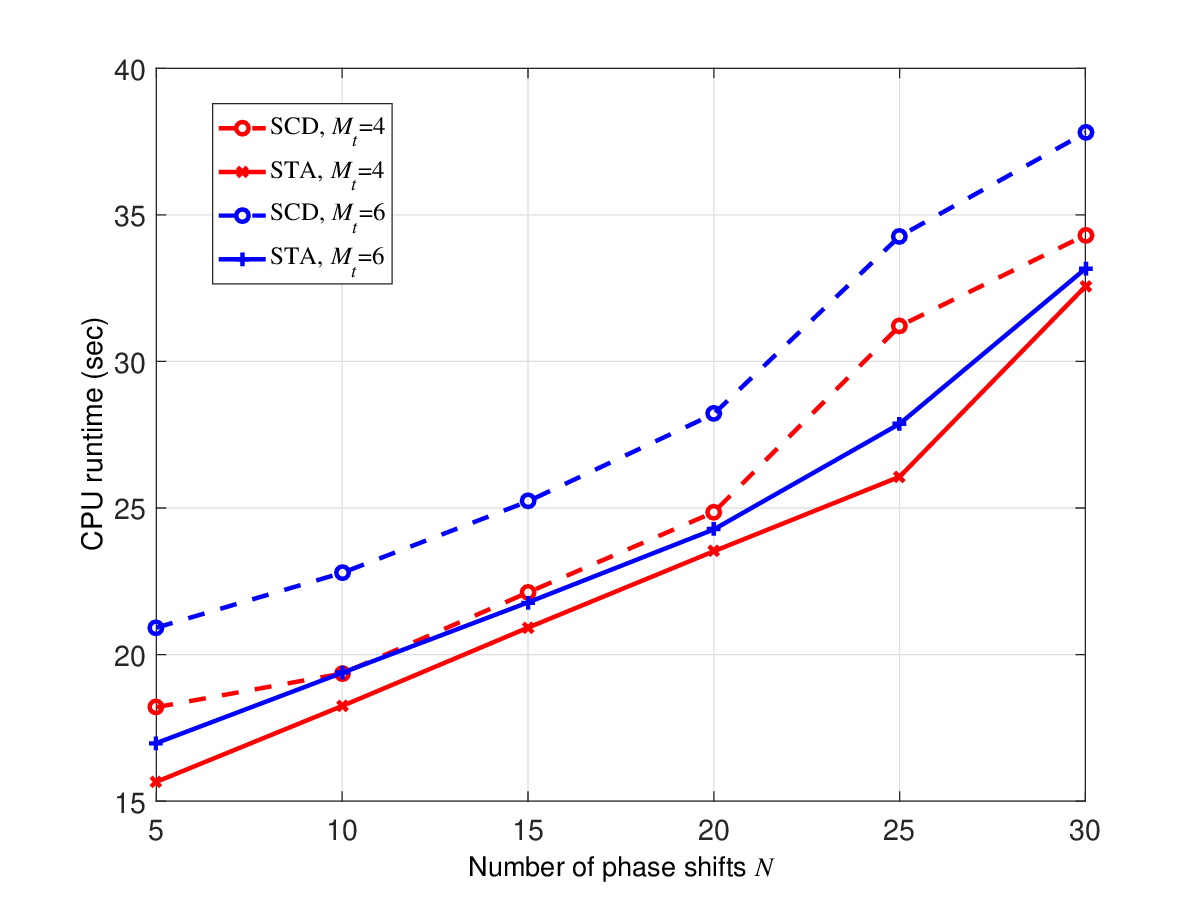}
\caption{CPU runtime versus $N$, with $r=1$, $\Gamma=-70$ dBm, $\beta=0.05$ and $\delta_{g}=0.05$.}
\label{runtime_N}
\end{minipage}
\end{figure}

\begin{figure}[htbp]
	\centering
\begin{minipage}[t]{0.48\textwidth}
	\centering
\includegraphics[width=3.1in]{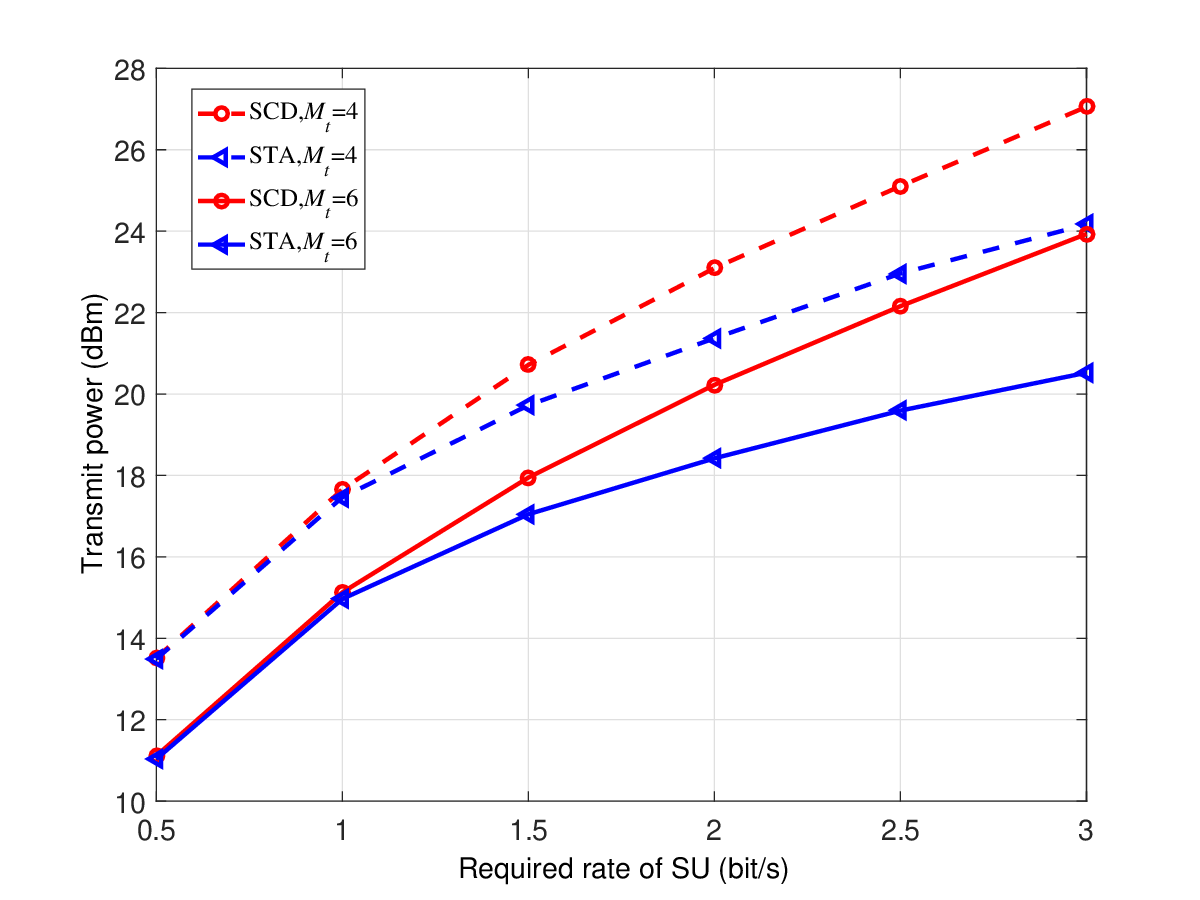}
\caption{Transmit power versus minimum required bit rate of SU, with $N=6$, $\Gamma=-80$ dBm, $\beta=0.05$ and $\delta_{g}=0.2$.}
\label{transpow_rate}
\end{minipage}
\begin{minipage}[t]{0.48\textwidth}
\centering
\includegraphics[width=3.1in]{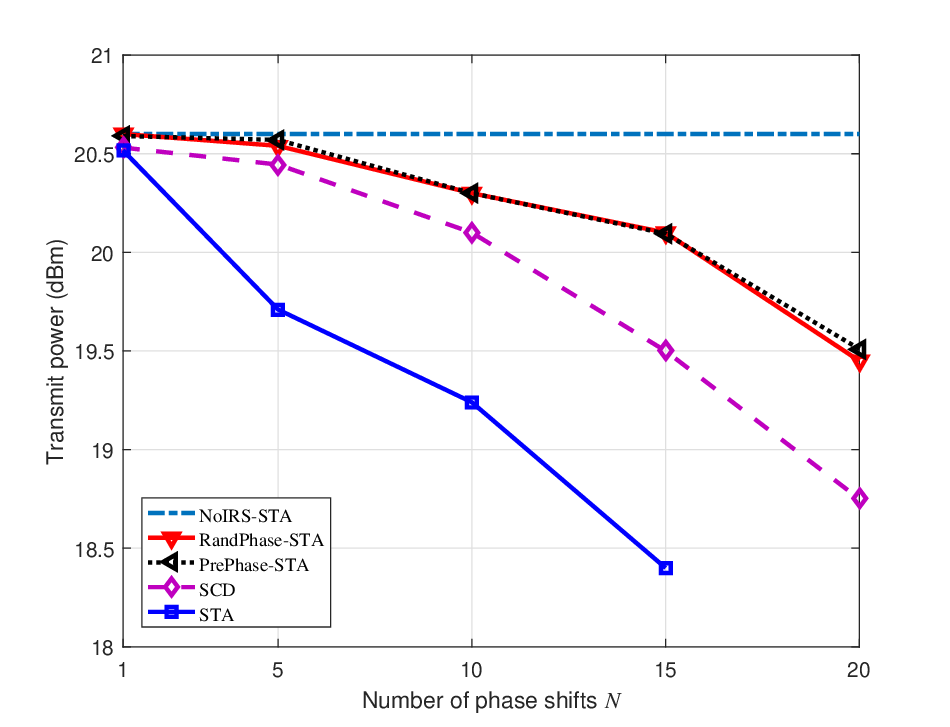}
\caption{Comparison with benchmarks with $r=2$, $M_t=6$, $\Gamma=-80$ dBm, $\beta=0.05$ and $\delta_{g}=0.1$.}
\label{transpow_N_compare}
\end{minipage}
\end{figure}

\par Consider the fact of the nonconvexity of problems $\mathcal{P}1$ and $\mathcal{P}2$,
different initial points may result in different locally optimal
solutions obtained by our proposed algorithms. By testing 30 randomly channel realizations, Fig. \ref{optimility_channelreal} illustrates the impact of the initializations on the performance of the proposed algorithms. The initializations of $\mathbf{W}$ and $\mathbf{\Phi}$ for SCD.alg and STA.alg
are determined by our proposed feasibility checking problems. SCD.alg-Exh and STA.alg-Exh refers to the best initial point of 1000 random initial points for each channel realization. It can be
seen that the minimum transmit power obtained by SCD.alg (STA.alg) is almost the
same as that of SCD.alg-Exh (STA.alg-Exh), which implies that our proposed algorithm is not sensitive to the initial points.
According to the target convergence criteria $\frac{\|\mathbf{W}^{(t+1)}\|_\mathrm{F}^2-\|\mathbf{W}^{(t)}\|_\mathrm{F}^2}{\|\mathbf{W}^{(t+1)}\|_\mathrm{F}^2}<\zeta$ which is set in Algorithm 1 and Algorithm 2, Fig. \ref{convergence_accuracy} shows that the higher the requirement of convergence accuracy, the smaller the optimal target value. The x-axis values from $-4$ to $-1$ represent the accuracy from $10^{-4}$ to $10^{-1}$ which means that the accuracy requirement decreases. Given the convergence accuracy $\zeta=10^{-4}$, Fig. \ref{convergence_iteration} shows that both the proposed algorithms can converge within six iterations. Compared with STA algorithm, the SCD algorithm always converges at a higher transmit power when the number of phase shifts ranges from 5 to 15. Moreover, from Fig. \ref{transpow_rate}, we can find that the transmit power increases with SUs' data rate requirements and the SCD algorithm yields a higher transmit power than STA algorithm when each SU's data rate is in a high value range for different numbers of the transmit antennas. This is due to the fact that the SCD scheme is the worst-case optimization which is more conservative than the statistical optimization and requires more transmit power to ensure that each SU's target rate requirement for the worst-case CSI error realization can be satisfied.

\par For the computational complexity, Fig. \ref{runtime_N} shows the average CPU running time versus the number of phase shifts $N$ for both proposed algorithms when $r=1$, $\Gamma=-70$ dBm, $\beta=0.05$ and $\delta_{g}=0.05$. The results are obtained by using a computer with a 1.99 GHz i7-8550U CPU and 8 GB RAM. The STA algorithm requires much less CPU running time than that required by the SCD algorithm. This is due to fact that there are some large-dimensional LMIs that increase the computational complexity of the SCD algorithm. With the increase of the number of $N$, all the algorithms need more running time to obtain the optimal solutions. Fig. \ref{convergence_iteration}, Fig. \ref{runtime_N} and Fig. \ref{transpow_rate} indicate that the STA algorithm outperforms the SCD algorithm both in terms of optimal value and computational complexity.
\subsection{Compared with Other Benchmarks}
\par Fig. \ref{transpow_N_compare} compares the SCD and STA schemes with three benchmark schemes, named as NoIRS-STA, RandPhase-STA and Prephase-STA. For the NoIRS-STA scheme, there is no IRS deployed in the system that can be regarded as the traditional CR system. For the RandPhase-STA scheme, the phase of each reflection element is randomly and uniformly generated between 0 and $2\pi$ for each subproblem of optimizing RED matrix. For the Prephase-STA scheme, all modulus values of the phase shifts are fixed to be one. In the above three benchmark schemes, TPC matrix is optimized by the same method as used in Algorithm 2. Our proposed SCD and STA schemes outperform these three benchmark schemes. The NoIRS scheme is the worst one that verifies the benefit brought by introducing the IRS into CR systems. The Prephase and RandPhase schemes without phase shifts optimization have almost the same performance and are worse than SCD and STA schemes.

\subsection{Feasibility and Objective Evaluations}
\begin{figure}[!t]
\centering
\subfigure[Feasibility rate]{
\centering
\includegraphics[scale=0.5]{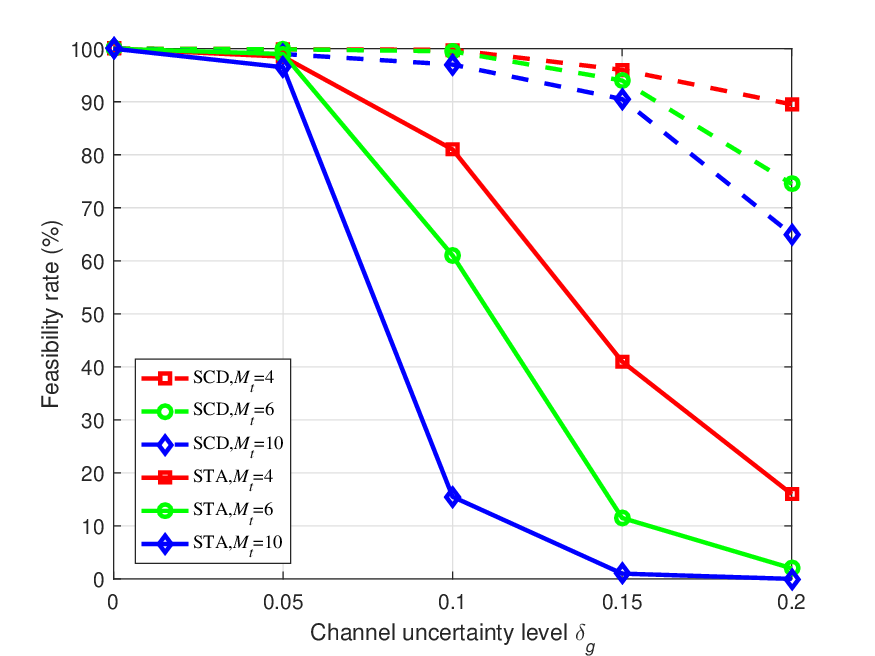}
\label{feasrate_delta}
}
\subfigure[Minimum transmit power]{
\centering
\includegraphics[scale=0.5]{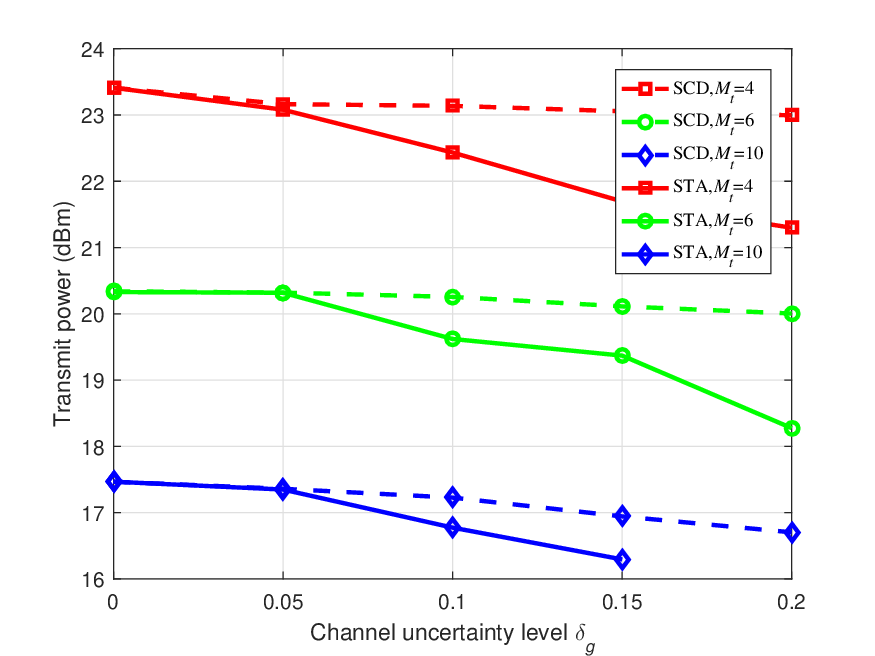}
\label{transpow_delta}
}
\caption{Feasibility rate and minimum transmit power versus channel uncertainty level $\delta_{g}$.}
\label{ft_delta}
\end{figure}
\par Fig. \ref{ft_delta} shows the feasibility rate and minimum transmit power versus the channel uncertainty level with various numbers of transmit antennas when $r=2$, $N=6$, $\Gamma=-80$ dBm and $\beta=0.05$. The feasibility rate is defined as the ratio of the number of feasible channel generations to the total number of channel generations, where the feasible channel generation means there exists a solution for the optimization problem under this channel generation. From Fig. \ref{feasrate_delta}, we can find that the feasibility rate decreases with the increase of both $\delta_g$ and the number of transmit antennas $M_t$. With the increase of $\delta_g$, the feasibility rate with large $M_t$ will reduce to zero more rapidly than that with small $M_t$. From Fig. \ref{transpow_delta}, the minimum transmit power will decrease with the increase of $M_t$, which is due to the following two reasons. The first one is that large $M_t$ improves the degrees of freedom which can be exploited to optimize the active beamforming at the ST. The second is that the PU-related channels become worse with the increase of $M_t$ and the IT imposed on PR becomes lower. The same results hold for $\delta_g$. With the increase of channel uncertainty, the PU-related channels become worse and more signal power is allocated to SUs, then the transmit power of ST can be reduced. Comparing Fig. \ref{transpow_delta} with Fig. \ref{feasrate_delta}, we can know that the values of the number of the transmit antennas and channel uncertainty level should be limited to achieve a good tradeoff between the feasibility rate and the ST's minimum transmit power in IRS-aided CR networks.
\par Note that, in Fig. \ref{transpow_delta}, there is no value for the blue solid line of STA when $M_t=10$. This is due to the fact that the corresponding feasibility rate shown in Fig. \ref{feasrate_delta} becomes 0 when channel uncertainty level $\delta_g$ is 0.2.
\begin{figure}[!t]
\centering
\subfigure[Feasibility rate]{
\centering
\includegraphics[scale=0.5]{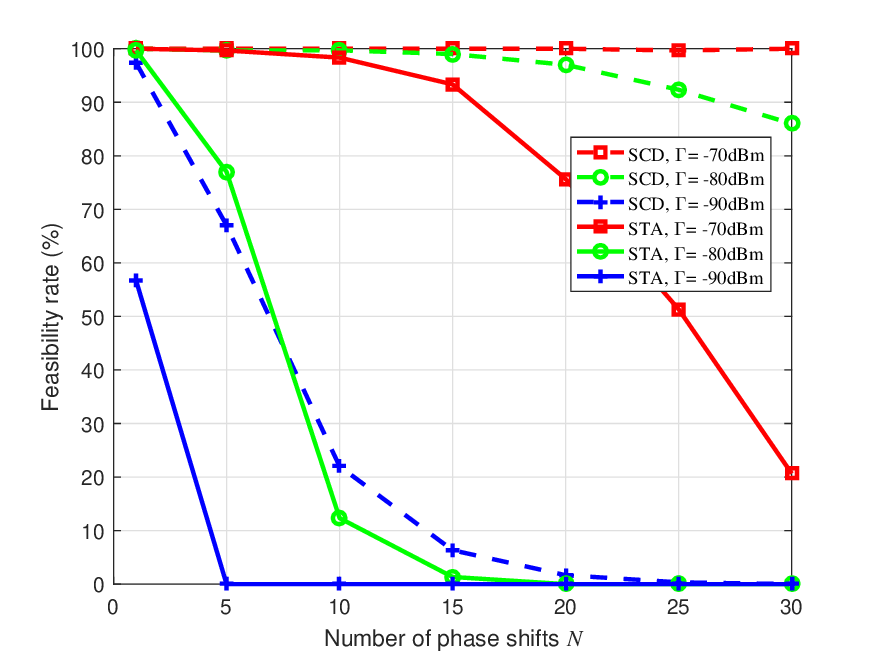}
\label{feasrate_N}
}
\subfigure[Transmit power]{
\centering
\includegraphics[scale=0.5]{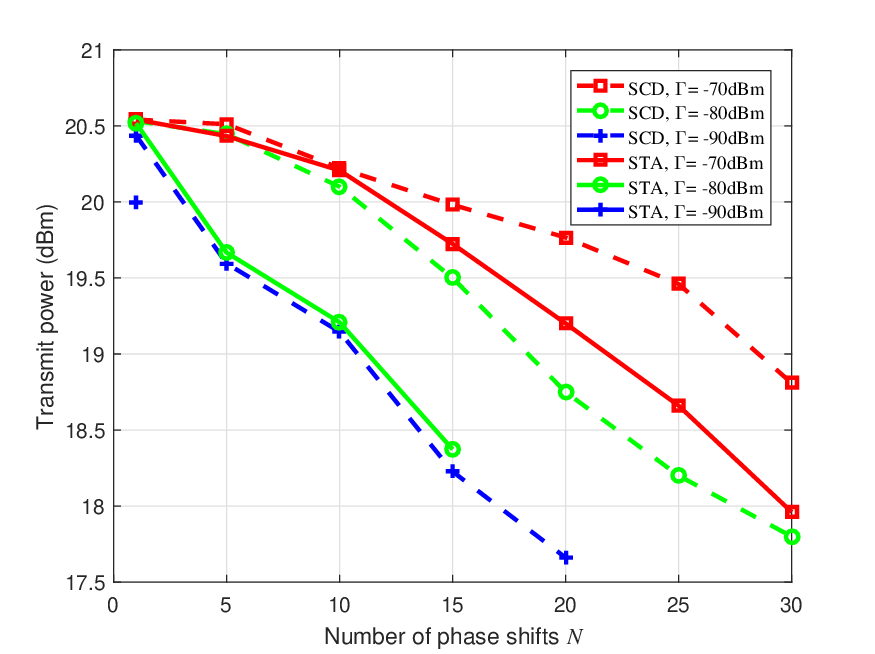}
\label{transpow_N}
}
\caption{Feasibility rate and minimum transmit power versus $N$.}
\label{ft_N}
\end{figure}
\par Fig. \ref{ft_N} shows the feasibility rate and the minimum transmit power versus the number of phase shifts $N$ for various values of IT threshold $\Gamma$ when $r=2$, $M_t=6$, $\beta=0.05$, $\delta_g=0.1$. From Fig. \ref{feasrate_N}, it is observed that both the feasibility rates of the SCD algorithm and the STA algorithm decrease with the increase of $N$. This is due to the fact that the cascaded ST-IRS-PR channel estimation error increases with $N$. Another phenomenon is that the feasibility rates of both the SCD and the STA algorithms decrease with $\Gamma$. The decrease of $\Gamma$ means the feasible space about IT limitation requirement shrinks. Fig. \ref{transpow_N} shows the minimum transmit power versus $N$ for various values of $\Gamma$. The ST's minimum transmit power decreases with the increase of $N$. This is due to the fact that increasing $N$ can enhance the reflective beamforming gain by optimizing the phase shift matrix. However, increasing $N$ can also increase the PU-related channel estimation error, which will reduce the feasibility rate. Hence, the number of phase shifts should be carefully chosen especially when the IT threshold is lower than $-80$ dBm. An extreme case can be found that when $\Gamma=-90$ dBm, the optimization problem is not feasible at almost all range of phase shifts from 5 to 30.
\subsection{Multi-PR Scenario Evaluations}\label{multiPRsim}
\begin{figure}[!t]
\centering
\subfigure[Feasibility rate]{
\centering
\includegraphics[scale=0.37]{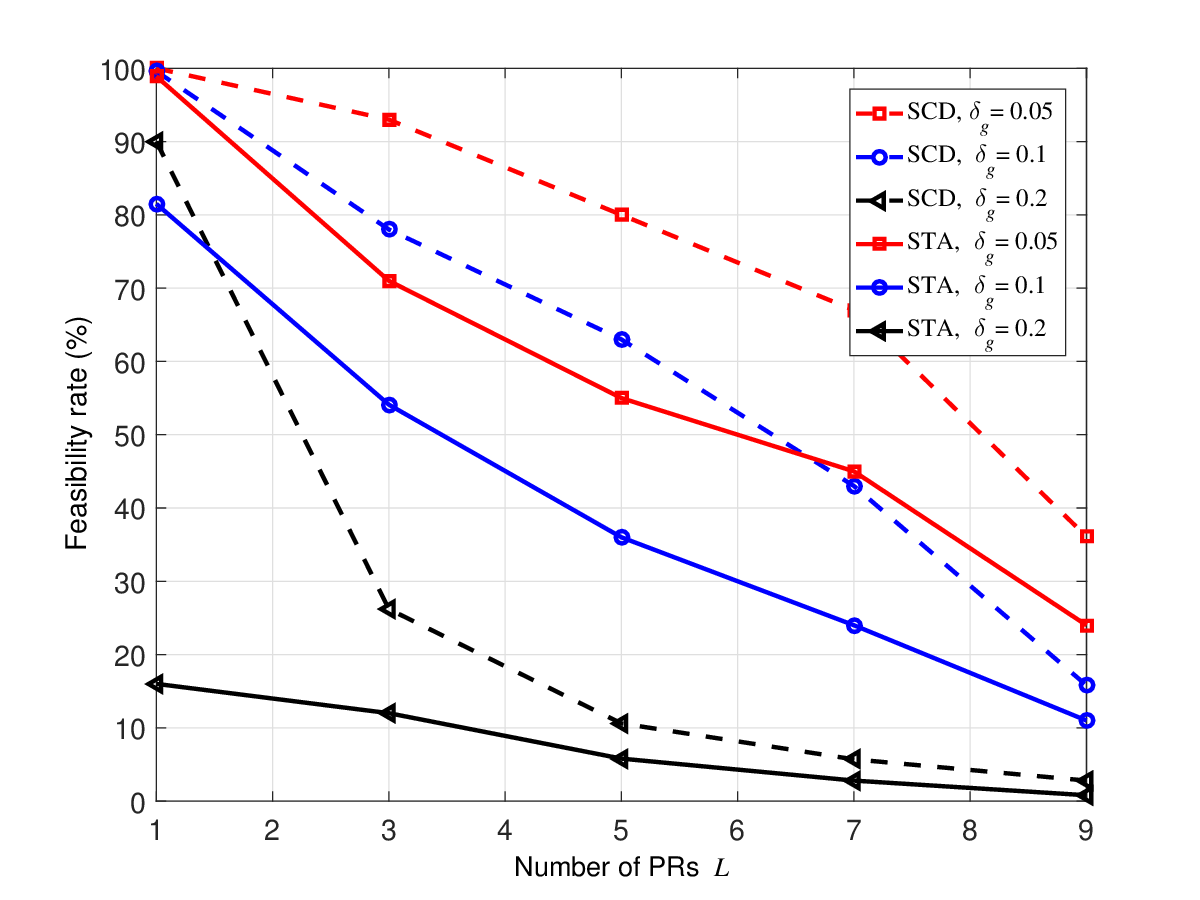}
\label{feasrate_multiPRs}
}
\subfigure[Transmit power]{
\centering
\includegraphics[scale=0.37]{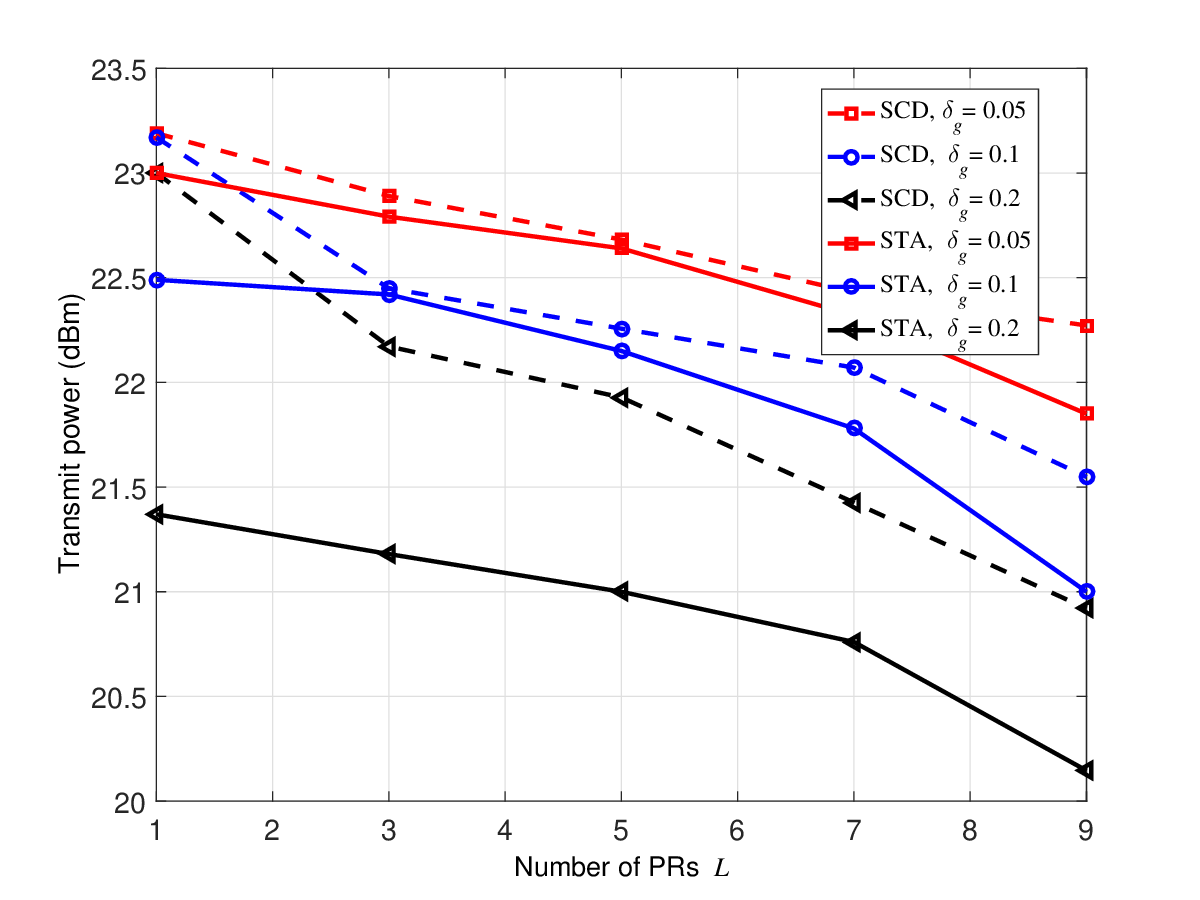}
\label{transpow_multiPRs}
}
\caption{Feasibility rate and minimum transmit power versus $L$.}
\label{ft_PRs}
\end{figure}
\par Fig. \ref{ft_PRs} shows the feasibility rate and the minimum transmit power versus the number of PRs $L$ for various values of channel uncertainty level $\delta_g$ when $r=2$, $M_t=4$, $\beta=0.05$, $N=6$. The PRs are randomly located in a cell region centered at origin with radius $200$ m and other scenario settings remain unchanged as shown in Fig. \ref{fig2_simu}. For PR $l$, $\forall l\in \mathcal{L}$, we set $\delta_{g_{d,l}}=\delta_{g_{r,l}}=\delta_g$ and $\Gamma_l=-80$ dBm. From Fig. \ref{feasrate_multiPRs}, it is observed that both the feasibility rates of the SCD algorithm and the STA algorithm decrease with the increase of the number of PRs and channel uncertainty level. This is due to the fact that the feasibility region shrinks with the increase of the number of constraints brought by increasing $L$ and $\delta_{g}$. Fig. \ref{transpow_multiPRs} shows the minimum transmit power versus $L$ for various values of $\delta_g$. For the same scheme, with the increase of $\delta_g$, the PU-related channels become worse and more signal power is allocated to SUs, then the transmit power of ST can be reduced. The ST's minimum transmit power decreases with the increase of $L$. The reason can be explained as follows. With the increase of the number of PRs, the number of PRs that are close to the ST will increase as well. In order to guarantee interference constraint imposed on PRs, the transmit power of ST should be reduced.

\subsection{Evaluations of Locations of PR and IRS}
\begin{figure}[!t]
	\centering
	\includegraphics[scale=0.52]{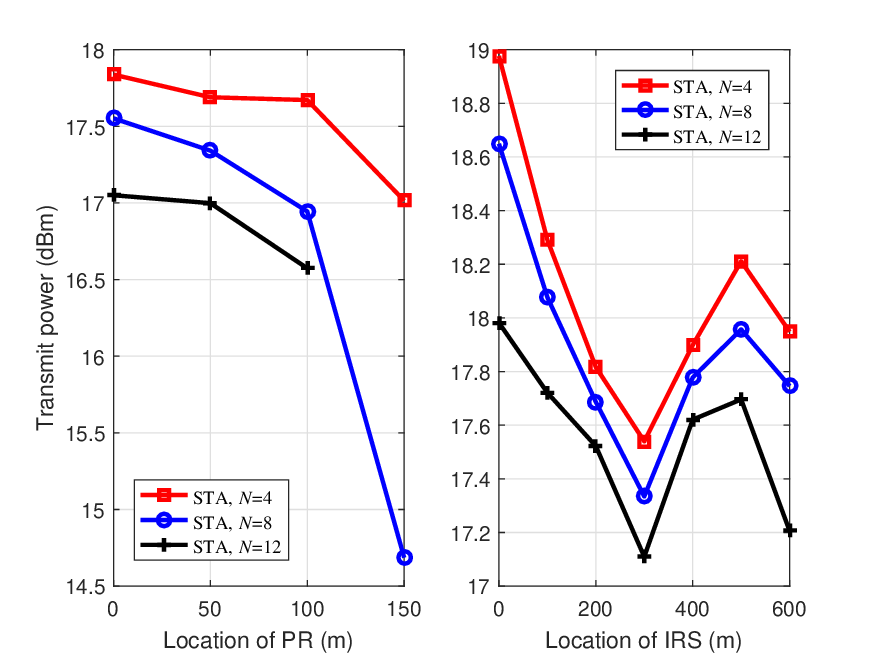}
	\caption{Impacts of PR and IRS locations on minimum transmit power.}
	\label{power_vs_PRandIRS_local}
\end{figure}
By moving PR from the original location (0 m, 0 m) to (150 m, 0 m) or IRS from (0 m, 30 m) to (600 m, 30 m) along the x-axis, respectively, we investigate the impact of PR location or IRS location on the performance of the network in Fig. \ref{power_vs_PRandIRS_local}, where we set $r=1$, $M_t=4$, $\beta=0.05$, $\Gamma=-70$ dBm and $\delta_g=0.05$. Fig. \ref{power_vs_PRandIRS_local} indicates that when PR is closer to ST, the minimum transmit power is smaller. This is because that the closer the PR is to the ST, the greater the possibility of the ST interfering with the PR. The ST has to decrease the transmit power. However, this will result in reducing the number of SUs allowed to access the channel. This is illustrated by the black line in first sub-figure where the optimization problem has solution only when the location of PR is less than 150 m. Another result indicated by the second sub-figure is that the location of IRS can affect the minimum transmit power of ST. When the IRS is deployed near the SU (ST or SR), such as at 300 m or 600 m, the transmit power can be lowest. This result can guide us where to deploy the IRS in IRS-aided CR systems.

\section{Conclusion}\label{section5}
In this paper, we have investigated two types of CSI error models for the PU-related channels in IRS-aided CR networks. Two schemes, i.e., SCD and STA were proposed to jointly optimize the TPC matrix and phase shift matrix. Simulation results show that if the estimated CSI error of the PU-related channel is large, the optimization problem has a higher probability to be infeasible. Even though the CSI error is small, the number of transmit antennas at the ST and the number of phase shifts of the IRS should be carefully chosen to balance the feasibility rate of the optimization problem and the total minimum transmit power. In order to achieve a certain feasibility rate for the SUs, the admission access control scheme will be another research direction in our future work.
\appendices
\section{Proof of Proposition \ref{proposition3} }\label{append4}
For the complex matrix $\triangle\mathbf{G}$ where the vector $\mathrm{vec}(\triangle\mathbf{G})$ is a Gaussian distribution vector, $\mathrm{vec}(\triangle\mathbf{G})$ can be normalized as
$\mathrm{vec}(\triangle\mathbf{G})\!\!=\!\!\mathbf{\Sigma}^{\frac{1}{2}}\mathbf{t}$,
where the vector follows the distribution of $\mathbf{t}\!\sim\!\! \mathcal{CN}\!(\!\mathbf{0},\mathbf{I}\!)$. As $\left\|\tilde{\boldsymbol{\phi}}^\mathrm{H}\triangle\mathbf{G}\mathbf{W}\right\|_2^2\!\!\leq\!\!\lambda_{max}(\mathbf{\Sigma})\!\left\|\tilde{\boldsymbol{\phi}}^\mathrm{H}\right\|_2^2
\left\|\mathbf{W}\right\|_\mathrm{F}^2\left\|\mathbf{t}\right\|_2^2$, the sufficient condition that the probability constraint in (\ref{eq_2_4}) holds is
\begin{equation}
\begin{split}
&\mathrm{Pr}\!\left\{\!(N\!+\!1)\lambda_{max}(\mathbf{\Sigma})\!\left\|\mathbf{W}\right\|_\mathrm{F}^2\left\|\mathbf{t}\right\|_2^2\!\leq\!\Gamma
\!\!-\!\!\left\|\tilde{\boldsymbol{\phi}}^\mathrm{H}\!\widehat{\mathbf{G}}\mathbf{W}\!\right\|_2^2 \right\}\\
&\!\geq\! 1\!\!-\!\!\beta\Leftrightarrow\mathrm{Pr}\left\{\left\|\mathbf{t}\right\|_2^2\leq \widetilde{\Gamma} \right\}\geq 1-\beta,
\end{split}
\end{equation}
where $\widetilde{\Gamma}=\frac{\Gamma
	 -\left\|\tilde{\boldsymbol{\phi}}^\mathrm{H}\widehat{\mathbf{G}}\mathbf{W}\right\|_2^2}{(N+1)\lambda_{max}(\mathbf{\Sigma})\left\|\mathbf{W}\right\|_\mathrm{F}^2}$.
\par Since $\mathbf{t}$ follows a zero-mean complex Gaussian distribution, $\left\|\mathbf{t}\right\|_2^2$ satisfies the chi-square distribution with $2(N+1)M_t$ degrees of freedom, i.e., $\left\|\mathbf{t}\right\|_2^2\sim\chi^2_{2(N+1)M_t}$. We define the cumulative distribution function $F_{2(N+1)M_t}(\widetilde{\Gamma})=\mathrm{Pr}\left\{\left\|\mathbf{t}\right\|_2^2\leq \widetilde{\Gamma} \right\}$ and the inverse cumulative distribution function $F_{2(N+1)M_t}^{-1}(\cdot)$. Then we have
\begin{equation}\label{eq_appendD2}
\begin{split}
\mathrm{Pr}\left\{\left\|\mathbf{t}\right\|_2^2\leq \widetilde{\Gamma} \right\}\geq 1-\beta
\Leftrightarrow&F_{2(N+1)M_t}^{-1}(1-\beta)\leq\widetilde{\Gamma}.
\end{split}
\end{equation}
By substituting $\widetilde{\Gamma}$ into (\ref{eq_appendD2}), we obtain the result in (\ref{eq_2_5}). In practice,
the inverse function of the central chi-square cumulative distribution function can
be evaluated directly or be stored in a lookup table in practical
implementation.
The proof is completed.

\section{Proof of Theorem \ref{theorem1} }\label{append3}
Denote the collection of $\{\mathbf{S}_k, \forall k \in \mathcal{K}\}$ as $\mathbf{S}$ and define $\boldsymbol{\lambda}=[\lambda_k, \forall k\in\mathcal{K}]$ and $\mu$ as the non-negative Lagrangian
	multipliers associated with SINR constraints (\ref{eq_problem3_1_b}) and IT constraint (\ref{eq_problem3_1_c}), respectively. Then, the Lagrangian function of Problem $\mathcal{P}2.1\_2$ is
	\begin{equation}
	\begin{split}
	 &\mathcal{L}(\mathbf{S},\boldsymbol\lambda,\mu)=\sum_{k\in\mathcal{K}}\mathrm{tr}(\mathbf{S}_k)-\sum_{k\in\mathcal{K}}\mathrm{tr}(\mathbf{S}_k\mathbf{Z}_k)\\
&+\!\!\sum_{k\in\mathcal{K}}\lambda_k\!\!\left(\!\!\gamma_k\!\!\left(\sum_{j\neq k,j\in\mathcal{K}}\!\!\mathrm{tr}(\mathbf{S}_j\widehat{\mathbf{H}}_{\phi,k})\!+\!\sigma_{s_k}^2\!\right)\!\!-\!\mathrm{tr}(\mathbf{S}_k\widehat{\mathbf{H}}_{\phi,k}\!)\!\!\right)\\
	&+\!\mu\!\left(\sum_{k=1}^K\!\mathrm{tr}\!\left(\mathbf{S}_k\!\left(\!\vartheta F_{2(N+1)M_t}^{-1}(\!1\!-\!\beta)\mathbf{I}\!+\!\widetilde{\mathbf{G}}_{\tilde{{\boldsymbol{\phi}}}}\right)\!\right)\!-\!\Gamma\right),
	\end{split}
	\end{equation}
	where $\{\mathbf{Z}_k\succeq\mathbf{0},\forall k\in\mathcal{K}\}$ denote the dual variable matrices associated with the semi-definite constraints on $\{\mathbf{S}_k,\forall k\in\mathcal{K}\}$.

\par As Problem $\mathcal{P}2.1\_2$ is a convex optimization problem, its globally optimal solution must satisfy its	first-order optimality condition as:
\begin{equation}
\frac{\partial\mathcal{L}(\mathbf{S},\boldsymbol{\lambda},\mu)}{\partial\mathcal{S}_k}=\mathbf{D}-\lambda_k\widetilde{\mathbf{H}}_{\phi,k}-\mathbf{Z}_k=0, \forall k\in\mathcal{K},
\end{equation}
where $\mathbf{D}$ is given by
\begin{equation}\label{eq_appendixB_0}
\mathbf{D}=\left(\mu\vartheta F_{2(N+1)M_t}^{-1}(1-\beta)+1\right)\mathbf{I}_{M_t}+\mu\widetilde{\mathbf{G}}_{\widetilde{\phi}}.
\end{equation}
Then, $\mathbf{Z}_k$ can be represented as $\mathbf{Z}_k=\mathbf{D}-\lambda_k\widetilde{\mathbf{H}}_{\phi,k}$.
According to the Karush-Kuhn-Tucker (KKT) condition, we have
\begin{equation}\label{eq_appendixB_1}
\lambda_k\!\!\left[\!\gamma_k\!\!\left(\sum_{j\neq k,j\in\mathcal{K}}\!\mathrm{tr}(\mathbf{S}_j\widehat{\mathbf{H}}_{\phi,k})\!+\!\sigma_{s_k}^2\!\right)\!\!-\!\mathrm{tr}(\mathbf{S}_k\widehat{\mathbf{H}}_{\phi,k})\right]\!\!=\!0, \forall k\!\in\!\mathcal{K},
\end{equation}
\begin{equation}\label{eq_appendixB_2}
\mathbf{S}_k\mathbf{Z}_k=\mathbf{0}, \forall k \in\mathcal{K}.
\end{equation}
Before proving the theorem, we first give the following lemma.
\begin{lemma} \label{lemma_appendixB}
	The optimal Lagrangian dual multipliers $\{\lambda_k, \forall k\}$ are positive, i.e., $\lambda_k>0, \forall k.$
\end{lemma}

\par
\emph{Proof}:This can be proved by using contradiction. Denote the optimal solution of Problem $\mathcal{P}2.1\_2$ as $\{\mathbf{S}_k^\star,\forall k\}$ and the corresponding Lagrangian multipliers as $\{\lambda_k^\star,\mu,\forall k\}$.
Assume there exists one $\lambda_l^\star$
that is zero. Then, according to KKT conditions in (\ref{eq_appendixB_1}), we have
\begin{equation}
\gamma_l\left(\sum_{j\neq l,j\in\mathcal{K}}\mathrm{tr}(\mathbf{S}_j^\star\widehat{\mathbf{H}}_{\phi,l})+\sigma_{s_l}^2\right)<\mathrm{tr}(\mathbf{S}_l^\star\widehat{\mathbf{H}}_{\phi,l}).
\end{equation}
Then, we can find a new precoding matrix $\mathbf{S}_l^o=\rho\mathbf{S}_l^\star$ with $0<\rho<1$ such that
\begin{equation}
\gamma_l\left(\sum_{j\neq l,j\in\mathcal{K}}\mathrm{tr}(\mathbf{S}_j^o\widehat{\mathbf{H}}_{\phi,l})+\sigma_{s_l}^2\right)\leq\mathrm{tr}(\mathbf{S}_l^o\widehat{\mathbf{H}}_{\phi,l})
\end{equation}
 holds. Then, we find a new feasible solution $\{\mathbf{S}_k^\star,k\neq l,\forall k, \mathbf{S}_l^o\}$ that yields a lower objective value than that of $\{\mathbf{S}_k^\star,\forall k\}$, which contradicts that $\{\mathbf{S}_k^\star,\forall k\}$ is the optimal solution.$\hfill \blacksquare$

\par
The first term of $\mathbf{D}$ in (\ref{eq_appendixB_0}) is an identity matrix multiplied by a constant. In addition, $\widetilde{\mathbf{G}}_{\widetilde{\phi}}$ is positive definite matrix and the Lagrange multiplier $\mu$ is not negative. As a result, $\mathbf{D}$ is positive definite matrix with $\mathrm{rank}(\mathbf{D})\!=\!M_t$. Since $\mathrm{rank}(\mathbf{Z}_k)\!\geq\!\mathrm{rank}(\mathbf{D})\!-\!\mathrm{rank}(\lambda_k\widehat{\mathbf{H}}_{\phi,k})$, we obtain $\mathrm{rank}(\mathbf{Z}_k)\!\geq\! M_t\!-\!1$, where we use the fact that $\mathrm{rank}(\lambda_k\widehat{\mathbf{H}}_{\phi,k})\!=\!1$ since $\widehat{\mathbf{H}}_{\mathbf{\Phi},k}\!=\!\widehat{\mathbf{h}}_{\mathbf{\Phi},k}\widehat{\mathbf{h}}_{\mathbf{\Phi},k}^\mathrm{H}$ and $\lambda_k$ is positive according to Lemma \ref{lemma_appendixB}. Furthermore, according to (\ref{eq_appendixB_2}), we have $\mathrm{rank}(\mathbf{S}_k)\!\leq\! M_t\!-\!\mathrm{rank}(\mathbf{Z}_k)$. Then, we have $\mathrm{rank}(\mathbf{S}_k)\!\leq\! 1$. Obviously, the
optimal precoder $\mathbf{S}_k$ is not zero matrix with zero-rank. Hence, the
optimal precoder $\mathbf{S}_k$ has rank one. The proof is completed.

\ifCLASSOPTIONcaptionsoff
  \newpage
\fi

\bibliographystyle{IEEEtran}
\bibliography{IRS_Robust_SRS_PR_MISO}



%
%




\end{document}